\documentclass{aa}

\usepackage{graphicx}
\usepackage{txfonts}
\usepackage{siunitx}
\usepackage{orcidlink}
\usepackage{xcolor}
\usepackage{placeins}
\usepackage{float}
\usepackage{hyperref}
\usepackage{letltxmacro}

\begin{document}

   \title{Active galactic nucleus activity in brightest cluster galaxies at intermediate redshifts in Sunyaev-Zel’dovich--selected clusters}
\titlerunning{AGN activity in brightest cluster galaxies at intermediate redshifts in SZ clusters}

   \author{N.~Hatamkhani\,\orcidlink{0000-0001-7075-0155} \inst{1,2}\fnmsep\thanks{\email{n.hatamkhani@gmail.com}}
          \and
          R.E. Skelton\,\orcidlink{0000-0001-7393-3336}
          \inst{1,2}
          \and
           J. Delhaize\,\orcidlink{0000-0002-6149-0846}
          \inst{2}
          \and
          M. Hilton\,\orcidlink{0000-0002-8490-8117}
          \inst{3,4}
          \and
          S.I. Loubser\,\orcidlink{0000-0002-3937-7126}
          \inst{5,6}
          \and
          N. Gupta\,\orcidlink{0000-0001-7547-4241}
          \inst{7}
          \and
          K. M. Mogotsi\,\orcidlink{0000-0002-5136-7983}
          \inst{1,2,8}
          }

   \institute{South African Astronomical Observatory, P.O. Box 9, Observatory, Cape Town 7935, South Africa 
         \and     
              Department of Astronomy, University of Cape Town, Private Bag X3, Rondebosch 7701, South Africa
         \and
             Wits Centre for Astrophysics, School of Physics, University of the Witwatersrand, Private Bag 3, Johannesburg 2050, South Africa
         \and
             School of Mathematics, Statistics \& Computer Science, University of KwaZulu-Natal, Westville Campus, Durban 4041, South Africa
         \and
             Centre for Space Research, North-West University, Potchefstroom 2520, South Africa
         \and
             National Institute for Theoretical and Computational Sciences (NITheCS), Potchefstroom 2520, South Africa
         \and
             Inter-University Centre for Astronomy and Astrophysics, Post Bag 4, Ganeshkhind, Pune 411 007, India
         \and
             Southern African Large Telescope (SALT), P.O. Box 9, Observatory 7935, Cape Town, South Africa
         }

\abstract
{Brightest cluster galaxies (BCGs) host some of the most powerful active galactic nuclei (AGNs) and play a central role in regulating the heating--cooling balance of the intracluster medium. While radio-mode feedback is well established in the local Universe, its prevalence and drivers at intermediate redshifts remain poorly constrained in representative cluster samples.}
{We investigated the incidence, accretion state, and host-galaxy properties of AGNs in BCGs residing in Sunyaev--Zel'dovich-selected clusters in the redshift range $0.3 < z < 0.8$, testing dependences on cosmic epoch, cluster mass, and cluster dynamical state.}
{We analysed 171 BCGs drawn from the Atacama Cosmology Telescope Sunyaev--Zel'dovich catalogue, combining Southern African Large Telescope spectroscopy, WISE mid-infrared photometry, and Rapid ASKAP Continuum Survey radio data with \textsc{X-CIGALE} spectral energy distribution modelling to derive stellar masses, star formation rates, and Eddington-scaled accretion rates.}
{Thirty-eight BCGs (22\%) host radio-loud AGNs. Only two ($\sim$5\%) are classified as high-excitation radio galaxies; the remainder are low-excitation systems. Most AGNs occupy low Eddington ratios ($-3 \lesssim \log_{10}\lambda_{\mathrm{Edd}} \lesssim -1$), consistent with predominantly radiatively inefficient accretion. Accretion efficiency increases with redshift, whereas only weak or statistically insignificant correlations are found with cluster mass and dynamical state. Radio luminosity is independent of cluster mass. Compared to field radio AGNs, BCG radio-loud AGNs exhibit similar radio powers but are hosted by galaxies that are, on average, $\sim0.8$ dex more massive.}
{BCG radio-loud AGNs at $0.3 < z < 0.8$ are predominantly low-excitation, radiatively inefficient systems, indicating that maintenance-mode accretion was already widespread in massive clusters ($M_{500c} \simeq 3.2 \times 10^{14}\,M_\odot$) by $z \sim 1$. The observed trends suggest that black hole accretion is regulated primarily by fuel availability and cosmic evolution rather than by global cluster properties, while the cluster environment mainly influences the characteristic host-galaxy mass scale.}

   \keywords{Brightest Cluster Galaxies, Active Galactic Nuclei, Galaxy Evolution, Optical Spectroscopy, Radio Continuum
                }

   \maketitle

\section{Introduction}

Brightest cluster galaxies (BCGs) are the most massive and luminous galaxies in the Universe and reside near the centres of galaxy clusters, close to the minimum of the gravitational potential. Their evolution is closely linked to both the assembly history of their host haloes and the thermodynamic state of the intracluster medium (ICM; \citealt{DeLucia2007,McNamara2007}). Multiwavelength observations have established that active galactic nucleus (AGN) feedback in BCGs plays a central role in regulating star formation and the heating--cooling balance of the ICM (\citealt{Fabian2012,Voit2015}). Mechanical energy from relativistic jets inflates X-ray cavities and drives shocks, offsetting radiative cooling and preventing runaway gas condensation in cluster cores (\citealt{Rafferty2006,HlavacekLarrondo2012,Ruppin2021}). AGN feedback is therefore a key ingredient in models of massive galaxy and cluster evolution.

Active galactic nucleus feedback is commonly described in terms of two accretion modes: a radiatively efficient `quasar' (or `radiative') mode and a kinetically dominated `radio' (or `maintenance') mode \citep[e.g.][]{Hardcastle2007,BestHeckman2012,HeckmanBest2014}. Observationally, these modes correspond to high-excitation and low-excitation radio galaxies (HERGs and LERGs; \citealt{BestHeckman2012}), which broadly trace radiatively efficient and inefficient accretion states, respectively. HERGs are typically associated with cold-gas accretion at relatively high Eddington ratios and strong optical emission lines, whereas LERGs are powered by radiatively inefficient accretion and predominantly release energy through relativistic jets \citep{Hardcastle2020,Arnaudova2025}. Although recent studies suggest a continuum of accretion states rather than strict bimodality (\citealt{Whittam2022,Arnaudova2025}), the distinction between HERGs and LERGs remains a useful observational framework for linking AGN populations to black hole fuelling and feedback.

At low redshifts ($z \lesssim 0.3$), BCGs are thought to be predominantly powered by radiatively inefficient maintenance-mode AGNs, reflecting the predominance of LERGs among massive radio galaxies (\citealt{BestHeckman2012,Sabater2019}). Whether this picture persists at earlier cosmic epochs remains uncertain. Higher-redshift BCGs exhibit enhanced star formation activity (\citealt{McDonald2016}), while radiatively efficient accretion and cold-gas reservoirs become increasingly common in the broader galaxy population (\citealt{Smolcic2009,Pracy2016,Tacconi2018,Kondapally2023,Deka2026}). Consistent with this trend, the fraction of BCGs that host AGNs identified using mid-infrared (MIR) colours in Sunyaev--Zel'dovich (SZ)-selected clusters increases strongly with redshift to $z\sim1.3$ (\citealt{Somboonpanyakul2022}), and \citet{HlavacekLarrondo2013} found evidence that radiatively efficient AGN activity becomes more prevalent at earlier epochs. However, it remains unclear whether these trends reflect a genuine shift in the balance between radiatively efficient and inefficient accretion. The intermediate-redshift interval $0.3<z<0.8$ therefore provides a key laboratory for testing the evolution of AGN fuelling and feedback in BCGs.

The role of the large-scale cluster environment in regulating AGN activity also remains uncertain. Cluster mass traces the depth of the gravitational potential and the overall gas reservoir, and the offset between the BCG and the cluster centre is commonly used as an indicator of a cluster's dynamical state and recent mergers. Such offsets have been linked to cool-core properties and to enhanced star formation in BCGs (\citealt{Sanderson2009,Hamer2012,Liu2025}), suggesting that cluster-scale processes influence the availability of cold gas and, consequently, AGN fuelling. However, it remains unclear whether these environmental factors regulate the level or mode of AGN activity once established.

Progress has been limited by both cluster selection effects and heterogeneous AGN diagnostics. Many previous studies have relied on X-ray or optically selected cluster samples, which preferentially identify cool-core or dynamically relaxed systems and may therefore bias inferred AGN demographics (\citealt{HlavacekLarrondo2012,HlavacekLarrondo2015}). For example, the incidence of radio-loud AGNs in BCGs rises to $\sim$70--90\% in strong cool-core clusters, compared to only $\sim$20--40\% in non-cool-core systems (\citealt{Best2007,Dunn2008}). In addition, many intermediate-redshift studies have focused on small targeted samples or radio-selected AGN populations that lack homogeneous host-cluster characterisation. As a result, the prevalence of radiatively efficient versus inefficient accretion in representative massive haloes remains poorly constrained.

Sunyaev--Zel'dovich surveys provide an approximately mass-limited census of galaxy clusters that is largely insensitive to cooling-core strength and only weakly dependent on redshift (\citealt{Planck2016,Hilton2021}), making them well suited for studying AGN demographics in representative massive haloes across cosmic time. The reliable classification of AGN accretion modes nevertheless requires multiwavelength diagnostics. Optical excitation indicators such as the equivalent width (EW) of the [O\,\textsc{iii}]~5007\,\AA\ emission line distinguish radiatively efficient from inefficient systems (\citealt{BestHeckman2012,Arnaudova2025}), MIR colours trace hot dust heated by luminous accretion (\citealt{Stern2012,Assef2018}), and radio continuum observations provide a direct probe of relativistic jets and kinetic AGN feedback (\citealt{Hardcastle2007,Cavagnolo2010,Hardcastle2020}). Combined with spectral energy distribution (SED) modelling, these diagnostics provide a comprehensive view of black hole growth and host-galaxy properties.

We investigated AGN activity in 171 BCGs residing in SZ-selected clusters from the Atacama Cosmology Telescope (ACT) Data Release 5 (DR5) catalogue \citep{Hilton2021} over the redshift range $0.3 < z < 0.8$. We combined optical spectroscopy from the Southern African Large Telescope \citep[SALT;][]{Buckley2006}, MIR photometry from the Wide-field Infrared Survey Explorer \citep[WISE;][]{Wright2010}, radio continuum observations from the Australian Square Kilometre Array Pathfinder \citep[ASKAP;][]{Johnston2008} obtained through the Rapid ASKAP Continuum Survey \citep[RACS;][]{Hale2021}, and multiwavelength SED fitting with \textsc{CIGALE} \citep{Boquien2019} to classify AGN accretion modes and derive host-galaxy properties.
The approximately mass-limited nature of the SZ-selected sample minimises biases related to cool-core strength and cluster dynamical state, enabling a cleaner assessment of how AGN activity depends on cosmic epoch, host-galaxy properties, and cluster environment. This dataset enabled us to address three key questions: (i) what fraction of AGN-hosting BCGs at intermediate redshifts are powered by radiatively efficient versus inefficient accretion, (ii) to what extent accretion properties depend on the cluster environment, including the cluster mass and dynamical state, compared to the cosmic epoch, and (iii) whether radio-loud AGNs in BCGs are intrinsically distinct from the general field population. By addressing these questions in a statistically uniform, approximately mass-limited cluster sample, we place new empirical constraints on the drivers of AGN activity and feedback in the most massive haloes since $z \sim 1$.

This paper is organised as follows. In Sect.~\ref{sec:data} we describe the cluster sample and the multiwavelength data used in this study. In Sect.~\ref{sec:sedfitting} we present the SED fitting procedure used to derive stellar masses, star formation rates (SFRs), and AGN contributions. Section~\ref{sec:agnclass} describes the optical, radio, and MIR diagnostics used to classify AGN activity and excitation states. The host-galaxy properties, radio characteristics, and accretion rates of the BCG population are presented in Sect.~\ref{sec:results}. In Sect.~\ref{sec:discussion} we discuss the implications of our results for AGN fuelling and feedback in cluster environments, and Sect.~\ref{sec:conclusion} summarises our main conclusions. All derived quantities assume a flat $\Lambda$ cold dark matter cosmology with $H_0 = 70$\,km\,s$^{-1}$\,Mpc$^{-1}$, $\Omega_{\rm M} = 0.3$, and $\Omega_\Lambda = 0.7$.

\section{Sample selection and data}
\label{sec:data}

\subsection{Cluster sample}
\label{sec:clustersample}

This study is based on galaxy clusters drawn from the ACT catalogue as part of the BCG Evolution with ACT, MeerKAT, and SALT (BEAMS\footnote{\url{https://astro.ukzn.ac.za/~beams/}}; PI: M. Hilton) programme. Clusters were identified via the thermal SZ effect \citep{Sunyaev1972}, which arises from inverse Compton scattering of cosmic microwave background photons by hot intracluster electrons. Because the SZ signal scales with the integrated electron pressure along the line of sight and is approximately redshift-independent, surveys selected via the SZ effect provide a selection that is closer to halo-mass selection than optical or X-ray surveys, with reduced sensitivity to cool-core strength.

\begin{figure}
    \centering
    \includegraphics[width=\columnwidth]{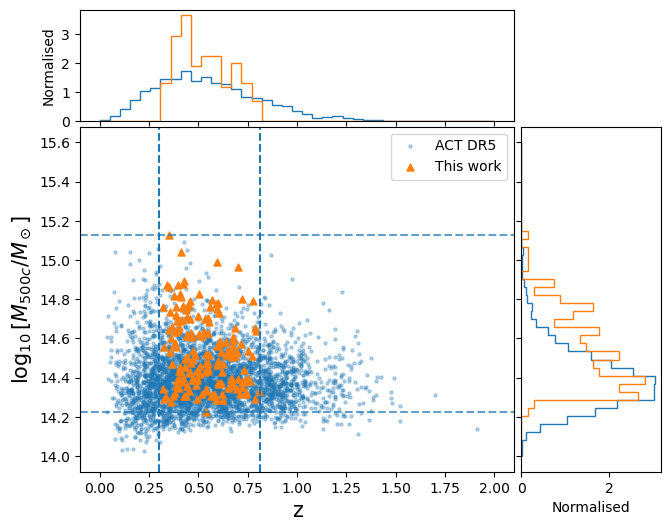}
    \caption{Redshift--mass distribution of the ACT DR5 clusters (blue points) and the BEAMS sample (orange triangles). Vertical dashed lines indicate the adopted redshift range, and horizontal dashed lines mark the cluster mass range of the BEAMS sample. The increasing mass threshold with redshift reflects the ACT selection function. Marginal histograms show the normalised redshift and mass distributions of both samples.}
    \label{fig:ACT_map}
\end{figure}

Our parent sample consists of 171 SZ-selected systems spanning the redshift range $0.3 < z < 0.8$. Clusters were first identified in the ACT DR4 maps \citep{Aiola2020} and subsequently confirmed in the final ACT DR5 catalogue \citep{Hilton2021}. We restricted our analysis to systems located within the footprint of the Dark Energy Survey \citep[DES;][]{Abbott2018} to ensure homogeneous optical photometry and reliable optical counterparts. Only clusters with signal-to-noise ratios (S/N) greater than five in the SZ detection significance were included, yielding an approximately mass-limited sample over the adopted redshift range.
The selected clusters span masses of $1.7 \times 10^{14} \lesssim M_{500c}/M_\odot \lesssim 1.4 \times 10^{15}$,\footnote{$M_{500c}$ is the mass enclosed within a radius corresponding to 500 times the critical density at the cluster redshift.} with a median value of $M_{500c} \simeq 3.2 \times 10^{14}\,M_\odot$. The sample is restricted to the redshift interval $0.3 < z < 0.8$, reflecting the selection criteria for SALT spectroscopic follow-up. As shown in Fig.~\ref{fig:ACT_map}, the ACT selection function introduces a mild redshift dependence, such that progressively more massive clusters dominate the sample at higher redshifts. Cluster masses are derived using an SZ--mass scaling relation calibrated against X-ray observations \citep{Arnaud2010}. Such calibrations are known to yield masses that are systematically lower than weak-lensing estimates for SZ-selected clusters by approximately $30\%$ \citep[e.g.][]{Robertson2024,Shirasaki2024}.
We adopted cluster redshifts from the ACT DR5 catalogue \citep{Hilton2021}, which combines optical photometric redshifts with spectroscopic confirmations from public surveys where available. The redshift--mass distribution of the parent ACT DR5 cluster catalogue and the subset that meets our BEAMS selection criteria is shown in Fig.~\ref{fig:ACT_map}.

The BCGs were identified following the BEAMS procedure \citep[see][]{Loubser2025}, selecting the most luminous extended galaxies in DES $r$-band imaging within $0.1\,R_{500}$ of the ACT SZ centroid. This search radius is sufficiently large to include moderately offset BCGs in dynamically disturbed systems, while minimising contamination from projected foreground galaxies and bright satellites. Candidate BCGs were further verified using the cluster red sequence in DES colour--magnitude diagrams and cross-checked against the DES photometric redshift catalogue. In cases of ambiguous optical morphology or projected companions, the BCG identification was confirmed by visual inspection. This procedure ensures a consistent BCG definition across the sample while avoiding strong bias against clusters with non-zero BCG--SZ offsets, which are later used as a diagnostic of cluster dynamical state. The resulting dataset forms the parent sample for the multiwavelength analysis presented in the following sections, integrating optical spectroscopy from SALT, MIR photometry from WISE, and radio data from RACS.

\subsection{Optical spectroscopy}
\label{sec:optical_spectroscopy}

Optical spectroscopy of the BCGs was obtained with the \textit{Robert Stobie} Spectrograph (RSS; \citealt{Burgh2003}) on the 10~m SALT \citep{Buckley2006}. Observations were carried out under three approved SALT programmes: the BEAMS Large Science Programmes (2019-1-LSP-001 and 2022-2-MLT-003; PI: M.~Hilton) and the follow-up programme 2024-1-SCI-031 (PI: N.~Hatamkhani). Together, these campaigns provide a homogeneous spectroscopic dataset for 109 BCGs drawn from the ACT cluster sample described in Sect.~\ref{sec:clustersample}.
The RSS was configured with the PG1300 grating for clusters at $z < 0.65$ and the PG0900 grating for higher-redshift targets, providing a typical rest-frame wavelength coverage of 3800--5100\,\AA\ and a spectral resolution of $\simeq 3\,\AA$. A slit width of $1\farcs25$ was used for all observations. Targets at $z < 0.65$ were observed in three 800~s integrations, while higher-redshift systems were observed over two visits ($2 \times 3 \times 800~$s) to achieve comparable S/N. Observations were obtained under dark conditions with seeing typically $<2\farcs0$.

Data reduction was performed using the \texttt{PySALT} package \citep{Crawford2010} for basic corrections and calibrations, and the \texttt{RSSMOSPipeline}\footnote{\url{https://github.com/mattyowl/RSSMOSPipeline}; documentation at \url{https://rssmospipeline.readthedocs.io/en/latest/}} \citep{Hilton2018} for flat-fielding, wavelength calibration, extraction, and stacking of one-dimensional spectra. Redshifts were determined from high signal-to-noise spectra through direct visual identification of prominent absorption and emission features, typically Ca H and K, H\,$\beta$, and [O\,\textsc{iii}] $5007$\,\AA. The redshift solutions were verified using the \texttt{RSSMOSPipeline} visualisation tool, which displays each observed spectrum overlaid with Sloan Digital Sky Survey (SDSS) spectral templates\footnote{\url{https://classic.sdss.org/dr5/algorithms/spectemplates/}} corresponding to different galaxy morphologies. An example reduced SALT RSS spectrum is shown in Fig.~\ref{fig:bcg_example_abs}.
For the subset of BCGs displaying emission lines in the initial pipeline-reduced spectra, the data were reprocessed using the Image Reduction and Analysis Facility (\texttt{IRAF}; \citealt{Tody1986,Tody1993}) to perform relative flux calibration. This secondary reduction used spectrophotometric standard stars observed within a few nights of the science frames. The calibrated spectra were then used to measure EW, providing the optical basis for the classification of AGNs (see Sect.~\ref{sec:opticalclass}).

We compared the spectroscopic redshifts of the BCGs with the adopted cluster redshifts provided in the ACT DR5 catalogue. The ACT DR5 cluster redshifts are heterogeneous in origin, including spectroscopic measurements from public spectroscopic surveys (PublicSpec), previous South Pole Telescope (SPT) and ACT cluster catalogues, and the literature (Lit), as well as photometric cluster-redshift estimates \citep{Hilton2021}. For the subset classified as PublicSpec, the cluster redshift is derived from available member-galaxy spectroscopy using an iterative biweight estimator rather than from the redshift of the BCG alone \citep{Hilton2021}. The rest-frame velocity offset was computed as $\Delta v = c \, (z_{\rm BCG} - z_{\rm cl})/(1 + z_{\rm cl})$. The initial distribution shows a small number of extreme velocity offsets ($|\Delta v| \gtrsim$ several $\times 10^{3}$ km s$^{-1}$). We examined these systems individually and found that they represent a small number of projected or otherwise complex systems rather than genuine BCG peculiar velocities or a general limitation of the ACT photometric redshifts. We therefore applied an iterative $3\sigma$ clipping procedure solely for this consistency check; this clipping was not used to define the scientific sample analysed in the remainder of the paper. After clipping, the velocity-offset distribution is tightly centred on zero, with a median of $\Delta v = 0$ km s$^{-1}$ and a dispersion of $\sigma = 462^{+40}_{-49}$ km s$^{-1}$, where the uncertainties were estimated using bootstrap resampling. This demonstrates the consistency between the SALT BCG redshifts and the adopted ACT cluster redshifts. Such a dispersion is physically plausible, as BCGs are known to exhibit peculiar velocities of a few hundred km s$^{-1}$ relative to the cluster rest frame, particularly in dynamically disturbed systems \citep[e.g.][]{Coziol2009,Lauer2014}.

\subsection{Optical photometry (DES)}

Optical photometry for all BCGs was obtained from the DES DR~1 \citep{Abbott2018}, which provides deep $griz$ imaging over approximately 5000~deg$^2$ of the southern sky. All clusters in the BEAMS sample lie within the DES footprint, ensuring homogeneous photometric coverage.

For each BCG, we extracted total magnitudes (\texttt{MAG\_AUTO}) and associated uncertainties from the DES~DR1 catalogue in the $g$, $r$, $i$, and $z$ bands. These magnitudes were corrected for Galactic extinction using the \citet{Schlegel1998} reddening maps recalibrated by \citet{Schlafly2011}, adopting a standard $R_V = 3.1$ extinction law. The DES photometry provides a consistent measure of the stellar continuum emission, serving as the optical input for the SED fitting described in Sect.~\ref{sec:sedfitting}.

\subsection{Mid-infrared data}
\label{sec:mirdata}

Mid-infrared photometry for the BCG sample was obtained from the Wide-field Infrared Survey Explorer (WISE; \citealt{Wright2010}) and the AllWISE data release \citep{Cutri2013}, which combines imaging from the cryogenic and post-cryogenic mission phases. The AllWISE catalogue provides full-sky coverage in four bands centred at 3.4, 4.6, 12, and 22~$\mu$m (hereafter $W1$, $W2$, $W3$, and $W4$), with angular resolutions of 6\arcsec, 6.5\arcsec, 6.5\arcsec, and 12\arcsec, respectively. These bands trace both the stellar continuum ($W1$, $W2$) and warm dust emission heated by star formation or AGN activity ($W3$, $W4$).

Cross-matching between the DES optical coordinates and the AllWISE catalogue was performed using a $3\arcsec$ radius, consistent with the typical WISE astrometric precision (e.g. \citealt{Jarrett2011}). We adopted the AllWISE profile-fit magnitudes (e.g. \texttt{w1mpro}) for all measurements and applied the standard quality filters (\texttt{cc\_flags}~=~`0000' and \texttt{ext\_flg}~$\leq$~1) to minimise contamination from image artefacts or extended-source confusion \citep[e.g.][]{Cluver2014,Yao2020}. MIR counterparts were identified for 160 of the 171 BCGs.
Flux densities were derived from the catalogue magnitudes using the zero-points and conversion factors of \citet{Jarrett2011}, and corrected for Galactic extinction following \citet{Yuan2013} with $E(B{-}V)$ values from the recalibration of \citet{Schlafly2011}. The S/N values from the AllWISE catalogue were used to assess measurement quality. Nearly all BCGs have $W1$ and $W2$ detections with ${\rm S/N}>3$ \citep[as recommended by][]{Cutri2013}, whereas the longer-wavelength $W3$ and $W4$ bands frequently exhibit lower S/N owing to their reduced sensitivity and the weak MIR emission of passive systems. All four WISE bands were retained for the SED fitting (Sect.~\ref{sec:sedfitting}), regardless of individual S/N values. The inclusion of all WISE bands ensures uniform photometric coverage across the sample and enables consistent estimation of both the stellar and AGN-heated dust components.

We also investigated the availability of far-infrared (FIR) data from public \textit{Herschel} surveys \citep{Pilbratt2010} for the BCG sample. Although $\sim50$ sources have catalogued counterparts, the majority of these measurements have low S/N, with flux uncertainties comparable to or exceeding the measured fluxes. Such measurements can nevertheless provide upper limits on the FIR emission and hence useful constraints on the obscured SFR. However, \textit{Herschel} coverage is available for only a minority of the sample, which would introduce heterogeneous photometric constraints across the SED-fitting analysis. We therefore elected not to include the \textit{Herschel} photometry in the main analysis and instead performed the SED fitting uniformly for all galaxies using the DES and WISE datasets.

As a sensitivity test, we repeated the SED fitting for a subset of BCGs with available spectral and photometric imaging receiver (SPIRE) observations from \textit{Herschel} at 250, 350, and 500~$\mu$m. The inclusion of FIR data generally resulted in lower fitted SFRs, although the FIR-inclusive SFR uncertainties are generally large and the two estimates are broadly consistent within their quoted uncertainties. Moreover, none of the FIR-inclusive fits satisfied the $\chi^2_{\rm r}<3$ criterion adopted for the main analysis (see Sect.~\ref{sec:sed_procedure}). We therefore do not regard these solutions as demonstrably superior fits or use them to revise the SFRs adopted in this work. The comparison nevertheless highlights an additional systematic uncertainty in the absolute SFR normalisation associated with the absence of FIR constraints.

\subsection{Radio data}
\label{sec:radio}

Radio continuum data for the BCG sample were obtained from RACS \citep{Hale2021}. The RACS survey provides near-all-sky coverage south of declination $+41^{\circ}$ at a central frequency of 887.5~MHz, with an instantaneous bandwidth of 288~MHz. The survey has a typical beam full width at half maximum of $\sim15\arcsec$ and a median rms noise level of 0.25--0.30~mJy~beam$^{-1}$. Sources in the RACS catalogue are detected at a threshold of $\gtrsim5\sigma$ significance. Cross-matching between the optical BCG coordinates and the RACS catalogue \citep{Hale2021} was performed within a $15\arcsec$ radius, chosen to encompass the RACS beam while allowing for positional offsets. Each potential match was visually inspected using the \textsc{Cube Analysis and Rendering Tool for Astronomy} (\textsc{CARTA}; \citealt{Comrie2021}), which provides interactive visualisation of radio continuum images and spectral cubes. This step was used to assess whether the radio emission was associated with the optical BCG and to identify contamination from nearby sources or imaging artefacts.

A total of 38 out of 171 BCGs (22.2\%) have a RACS source within the search radius. Among these, 25 were visually confirmed as secure counterparts where the radio emission is clearly centred on the optical BCG. For the remaining 13 cases, the radio emission was spatially blended or offset from the optical position, making source attribution uncertain within the RACS beam. To avoid overestimating their radio emission, the measured integrated flux densities for these ambiguous sources were conservatively treated as upper limits. No flux densities were assigned to the remaining non-detections. An example of the radio--optical alignment for a secure counterpart is shown in Fig.~\ref{DES_racs}.

We note that an additional 15 BCGs show radio counterparts in deeper MeerKAT public surveys, such as the MeerKAT Galaxy Clusters Legacy Survey \citep{knowles2022} and the MeerKAT Absorption Line Survey \citep{Deka2024}. However, these data were not included in the present analysis in order to maintain a homogeneous radio dataset with a well-defined selection function. Combining surveys with different observing frequencies, sensitivities, and angular resolutions would introduce non-uniform detection limits and complicate the interpretation of AGN incidence and radio luminosity distributions.

For secure RACS counterparts, the 887~MHz flux densities were converted to rest-frame 1.4~GHz radio luminosities assuming a power-law spectrum of the form $S_{\nu} \propto \nu^{\alpha}$ with spectral index $\alpha = -0.75$, typical for BCG radio sources \citep{HlavacekLarrondo2013,Giacintucci2014}. The luminosities were computed as

\begin{equation}
    L_{1.4} = 4\pi D_{L}^{2}\, S_{887}\,
    \left(\frac{1.4\,\mathrm{GHz}}{0.887\,\mathrm{GHz}}\right)^{\alpha}\,
    (1+z)^{-(1+\alpha)},
\end{equation}

\noindent where $D_L$ is the luminosity distance (in metres) and $S_{887}$ is the observed flux density at 887~MHz. Flux densities are converted to SI units assuming $1~\mathrm{Jy} = 10^{-26}~\mathrm{W\,m^{-2}\,Hz^{-1}}$.

As shown in Fig.~\ref{fig:redshift_dist}, the redshift distribution of radio-loud BCG AGNs closely follows that of the parent sample. A two-sample Kolmogorov-Smirnov (KS) test yields $D = 0.072$ and $p = 0.995$, indicating no statistically significant difference between the two distributions. This provides no evidence for a strong redshift-dependent difference in the observed radio-detection fraction over the adopted redshift range.

\section{Spectral energy distribution fitting}
\label{sec:sedfitting}

\subsection{Method and models}
\label{sec:sed_method}

\begin{table*}[h!]
\caption{SED-fitting modules and parameter ranges adopted in \textsc{X-CIGALE}.}
\label{tab:sedgrid}
\centering
\begin{tabular}{ll}
\hline\hline
Component & Parameter grid  \\
\hline
Star formation history & Delayed-$\tau$ model + recent burst (max duration 50~Myr) \\
Age of main population (Myr) & 1000, 2000, 4000, 7000, 10000 \\
$e$-folding time $\tau$ (Myr) & 250, 500, 1000, 2000, 5000 \\
Stellar population & \citet{Bruzual2003} SSP; IMF = \citet{Chabrier2003}; $Z=0.008, 0.02, 0.05$ \\
Dust attenuation & \citet{Charlot2000} law; reddening $E(B{-}V)=0.0$--0.9 \\
Dust emission & \citet{Dale2014} templates; $\alpha = 0.125$--3.0 (slope in $dM_{dust}\propto U^{-\alpha}dU$) \\
AGN module (SKIRTOR) & $\tau_{9.7} = 3, 7, 11$; viewing angles = 40$^\circ$, 70$^\circ$, 90$^\circ$; $\mathrm{frac_{AGN}} = 0.01$--0.9 \\
 & Torus radial ratio $r_\mathrm{max}/r_\mathrm{min} = 20$; torus density radial and angular parameters $p = q = 1$ \\
 & Polar dust: $E(B{-}V)=0.03$, $T=100$~K, emissivity = 1.6, SMC law \\
\hline
\end{tabular}
\tablefoot{For the definitions of the various parameters, see Sect.~\ref{sec:sedfitting}.}
\end{table*}

The SEDs of the BCGs were modelled with the \textsc{X-CIGALE} code \citep{Boquien2019,Yang2020}, which combines stellar, dust, and AGN emission within an energy-balance framework. This approach self-consistently links the energy absorbed by dust in the ultraviolet--optical regime to its re-emission in the infrared, allowing the simultaneous determination of host-galaxy and AGN properties.

The adopted configuration follows \citet{Mountrichas2024}, with adjustments tailored to the BCG sample and the available photometric coverage from DES and WISE. Stellar emission was modelled using the \citet{Bruzual2003} single stellar population (SSP) templates with a \citet{Chabrier2003} initial mass function (IMF). We adopted a grid of stellar metallicities ($Z = 0.008, 0.02, 0.05$), spanning moderately sub-solar to super-solar values. Although BCGs are generally metal-rich systems with typical stellar metallicities of $\gtrsim 1$--$2\,Z_{\odot}$ \citep[e.g.][]{Loubser2009,Lidman2012,Bellstedt2016,Loubser2025}, the inclusion of a broader metallicity grid avoids imposing overly restrictive priors during the SED fitting. At the high-metallicity regime relevant for BCGs, stellar population properties are only weakly sensitive to metallicity variations, such that the impact on derived parameters, such as stellar masses, is expected to be modest ($\lesssim 0.1$--$0.15$ dex).

A delayed star formation history of the form $\mathrm{SFR}(t)\propto t\,\exp(-t/\tau_{\mathrm{main}})$ was adopted, including an optional 50~Myr burst component to account for possible residual or recent star formation. Nebular emission, including lines and continuum, was included consistently with the ionising photon budget to ensure accurate modelling of systems exhibiting weak line emission. The parameter grid spans $\tau_{\mathrm{main}}=250$--5000~Myr and $t_{\mathrm{main}}=1000$--10\,000~Myr, covering the range of formation timescales considered for massive cluster centrals at $0.3<z<0.8$ \citep[e.g.][]{Thomas2005,DeLucia2007,Lidman2012,McDermid2015}.

Dust attenuation followed the two-component prescription of \citet{Charlot2000}, while dust re-emission from starlight heating was modelled using the \citet{Dale2014} templates. The AGN component was represented by the clumpy-torus \textsc{SKIRTOR} library \citep{Stalevski2012,Stalevski2016}, which accounts for both type-1 and type-2 orientations and anisotropic infrared emission. The free parameters and their adopted ranges are summarised in Table~\ref{tab:sedgrid}.

\subsection{Photometric inputs and fitting procedure}
\label{sec:sed_procedure}

The SEDs were constructed using optical photometry from DES ($g$, $r$, $i$, $z$) and MIR photometry from AllWISE. These bands provide leverage on the stellar continuum and the MIR dust emission associated with both star formation and AGN heating. To mitigate the effects of heterogeneous photometric errors, we imposed a minimum relative uncertainty of 10\% on all fluxes, preventing individual bands with underestimated errors from dominating the likelihood function. The SED fitting analysis was performed for 160 BCGs with complete DES and WISE photometric coverage. All redshifts were fixed during the fitting. Of the 160 galaxies, 106 have spectroscopic redshifts from our SALT observations (Sect.~\ref{sec:optical_spectroscopy}), corresponding to the subset of the 109 spectroscopic BCGs that are included in the SED-fitting sample.
For the remaining 54 galaxies, we adopted the ACT DR5 cluster redshifts described in Sect.~\ref{sec:clustersample}. These comprise 37 spectroscopic and 17 photometric cluster redshifts. The spectroscopic measurements originate from PublicSpec, SPT, and ACT, while the photometric measurements originate from redMaPPer, zCluster, and SPT. Here, redMaPPer is the optical cluster finder introduced by \citet{Rykoff2014,Rykoff2016}, while zCluster is the photometric cluster-redshift algorithm described by \citet{Hilton2018}. Individual uncertainties are provided for the photometric redshifts in the ACT DR5 catalogue; for the 17 photometric redshifts used here, these range from 0.005 to 0.035. All adopted redshifts were fixed during the SED fitting.

\textsc{X-CIGALE} was executed in Bayesian mode, yielding both the minimum $\chi^{2}$ best-fit model and the likelihood-weighted mean values (\texttt{bayes.*}) of all physical parameters. A total of 148 out of 160 galaxies ($\simeq93\,\%$) satisfy a reduced $\chi^{2}_{\mathrm{r}} < 3$, indicating that the adopted model grid provides an overall acceptable representation of the observed fluxes. To ensure internal consistency, we further required the Bayesian and best-fit estimates of $M_{\star}$ and SFR to agree within a factor of five ($|\Delta \log X| < 0.7$ dex), following common practice in \textsc{CIGALE} analyses \citep[e.g.][]{Ciesla2015,Masoura2018,Mountrichas2022a}. Applying this additional criterion reduces the sample to 133 of 160 sources ($\simeq82\,\%$), which form the robust subsample used throughout the remainder of the analysis. The final subsample has a median reduced $\chi^{2}_{\mathrm{r}} \simeq 0.6$, indicating generally high-quality fits across the sample. For each retained galaxy, we adopted the Bayesian stellar mass (\texttt{bayes.stellar.m\_star}), SFR (\texttt{bayes.sfh.sfr}), and AGN fractional contribution (\texttt{bayes.agn.fracAGN}) as fiducial values.

\subsection{Uncertainties and reliability}
\label{sec:uncertainties}

Uncertainties were derived from the Bayesian posterior distributions computed by \textsc{X-CIGALE}. Typical uncertainties are $\sim0.13$ dex in stellar mass and $\sim0.4$ dex in SFR. Mock catalogue tests indicate that these parameters are recovered without significant systematic bias. Full details are provided in Appendix~\ref{appendix:SED-uncer}.

As an external sanity check, we compared our SED-derived properties for the most extreme system in the sample, the BCG Phoenix~A hosted by the Phoenix cluster (SPT-CL~J2344$-$4243), with values reported in the literature. Phoenix~A hosts a powerful radiatively efficient AGN and exhibits exceptionally high SFRs, making it a useful test of SED fitting in the presence of strong AGNs and dust emission. The corresponding SED fit is shown in Fig.~\ref{cigale_examp}. Our \textsc{X-CIGALE} modelling yields $\log(M_{\star}/M_{\odot}) = 11.90 \pm 0.15$ and $\log(\mathrm{SFR}/M_{\odot}\,\mathrm{yr}^{-1}) = 3.06 \pm 0.28$, corresponding to $\mathrm{SFR} \simeq 1.1\times10^{3}\,M_{\odot}\,\mathrm{yr}^{-1}$. Previous studies report SFRs of $\sim500$--800\,$M_{\odot}\,\mathrm{yr}^{-1}$ with uncertainties of $\sim0.1$--0.2 dex \citep[e.g.][]{McDonald2012,McDonald2015}, while more recent JWST observations find SFRs of $\sim740$--1340\,$M_{\odot}\,\mathrm{yr}^{-1}$ depending on the averaging timescale \citep{Reefe2025}. Stellar masses are typically $\log(M_{\star}/M_{\odot}) \sim 11.5$--11.7 with systematic uncertainties of $\sim0.3$ dex \citep[e.g.][]{Tozzi2015,Russell2017}.
Given the absence of FIR constraints and the strong AGN contribution to the MIR emission, our SFR estimate is higher than, but broadly consistent with, literature values within the expected systematic uncertainties. More generally, the SFRs derived for the BEAMS sample should be interpreted with some caution. Nevertheless, comparison with literature values indicates that the SED-derived stellar masses and SFRs are broadly consistent with independent estimates, even for systems with strong AGN contamination. For the remainder of the BEAMS BCG sample, which is less extreme than Phoenix~A, systematic uncertainties associated with AGN contamination are expected to be smaller.

\section{AGN classification}
\label{sec:agnclass}

\subsection{Optical classification}
\label{sec:opticalclass}

Optical emission lines provide one of the most direct tracers of nuclear activity in galaxies. In particular, the [O\,\textsc{iii}]~5007\,\AA\ line is widely used as a tracer of AGN photoionisation and radiative accretion activity \citep[e.g.][]{Kauffmann2003,Heckman2004,Kewley2006}. The presence and strength of this line reflect the ionisation state of the gas surrounding the central engine, with larger EWs generally corresponding to higher excitation and, by extension, higher accretion rates.

For each BCG in our spectroscopic sample with detected emission lines, we measured the EW of [O\,\textsc{iii}]~$5007$\,\AA\ in the SALT spectra using direct Gaussian fitting after local continuum subtraction. The EW uncertainties were estimated via Monte Carlo resampling of the continuum noise. The median continuum S/N is $\sim6$ per pixel, resulting in typical EW uncertainties of $\sim0.3$--$1.5$\,\AA, depending on line strength. Systems with detectable [O\,\textsc{iii}] emission are defined as those with $\mathrm{EW}/\sigma_{\mathrm{EW}} > 2$ and are classified as emission-line systems, while those lacking significant emission are considered passive.

Among the 109 SALT spectra, four BCGs show emission-line features, corresponding to an emission-line fraction of $\sim3.6\%$ (see Sect.~\ref{sec:herg_lerg} for details). The majority of the sample exhibits purely absorption-line spectra characteristic of old stellar populations, with no detectable nebular features. This low fraction of emission-line systems is consistent with previous optical studies of BCGs at comparable redshifts \citep[e.g.][]{Edwards2007,Loubser2018}, indicating that most BCGs lack strong optical signatures of radiatively efficient AGN activity in cluster centres. Although the SALT spectroscopic subsample broadly spans the same redshift and cluster-mass range as the full BEAMS sample, we caution that the measured emission-line fraction formally applies only to the spectroscopically observed subset.

\subsection{Radio classification}
\label{sec:radioclass}

Radio continuum emission provides a direct tracer of mechanical feedback from AGNs, arising from synchrotron radiation produced by relativistic jets and lobes. At low radio luminosities, however, radio emission can also originate from star formation, necessitating a diagnostic that can robustly separate AGN-dominated systems from purely star-forming galaxies. To classify the radio properties of the BEAMS BCGs, we adopted a method based on the infrared--radio correlation (IRRC).

\subsubsection{Infrared--radio correlation}
\label{sec:irrc}

The IRRC provides a well-established method for distinguishing radio emission powered by star formation from that associated with AGN activity. In star-forming galaxies, infrared emission traces dust heated by young stellar populations, while radio emission primarily arises from synchrotron radiation associated with supernova remnants. Galaxies that host radio-loud AGNs frequently deviate from this relation, exhibiting excess radio emission produced by relativistic jets and lobes \citep[e.g.][]{delhaize2017,delvecchio2021}.

We quantified deviations from the IRRC using the parameter
\begin{equation}
q_{\mathrm{IR}} =
\log_{10}\left(\frac{L_{\mathrm{IR}}}{3.75 \times
10^{12}\,\mathrm{W}}\right)
-
\log_{10}\left(\frac{L_{1.4}}{\mathrm{W\,Hz^{-1}}}\right),
\end{equation}
where $L_{\mathrm{IR}}$ is the total infrared luminosity integrated over 8--1000~$\mu$m and $L_{1.4}$ is the rest-frame 1.4~GHz radio luminosity. Infrared luminosities were taken from the Bayesian dust luminosity (\texttt{bayes.dust.luminosity}) returned by \textsc{X-CIGALE} (Sect.~\ref{sec:sedfitting}), which represents the total energy re-emitted by dust heated by both stellar populations and any AGN component. We note that inclusion of an AGN dust contribution would increase $L_{\mathrm{IR}}$ and therefore bias $q_{\mathrm{IR}}$ towards higher values, making our radio-excess classification conservative. Radio luminosities were derived from the RACS 887~MHz flux densities and converted to 1.4~GHz assuming a spectral index of $\alpha=-0.75$ (Sect.~\ref{sec:radio}).
We identified radio-excess sources using the stellar mass- and redshift-dependent IRRC derived by \citet{delvecchio2021}, as implemented by \citet{Whittam2022}. Following these works, a source is classified as radio-excess if its $q_{\mathrm{IR}}$ lies more than 0.43~dex below the best-fit IRRC (their Eq.~5), corresponding to a $2\sigma$ deviation, where $\sigma$ is the intrinsic scatter of the relation. The expected value of $q_{\mathrm{IR}}$ as a function of stellar mass and redshift is given by

\begin{equation}
\begin{aligned}
q_{\mathrm{IR}}(M_{\star}, z) =\;&
(2.646 \pm 0.024)\,(1+z)^{-0.023 \pm 0.008} \\
&- (0.148 \pm 0.013)\,[\log(M_{\star}/M_{\odot}) - 10],
\end{aligned}
\label{eq:bestq}
\end{equation}
where the first term describes the redshift evolution and the second term accounts for the dependence on stellar mass. \citet{delvecchio2021} demonstrated that this $2\sigma$ threshold provides an optimal compromise between completeness and contamination, with misclassification of star-forming galaxies limited to $\sim3$--$4\%$.
\begin{figure}
    \centering
    \includegraphics[width=0.8\columnwidth]{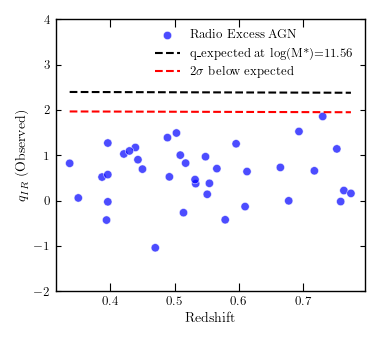}
    \includegraphics[width=0.8\columnwidth]{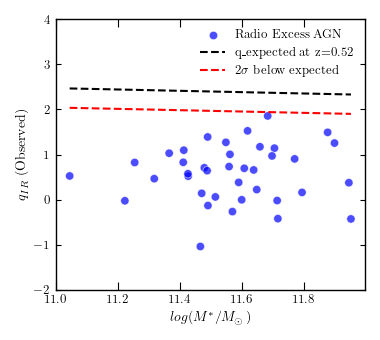}
    \caption{IRRC parameter ($q_{\mathrm{IR}}$) for the radio-excess AGNs. Blue points denote sources lying more than $2\sigma$ below the expected IRRC. The dashed black line shows the stellar mass- and redshift-dependent relation of \citet{delvecchio2021} (Eq.~\ref{eq:bestq}), while the dashed red line marks the corresponding $2\sigma$ threshold (0.43 dex below the relation). Top: $q_{\mathrm{IR}}$ versus redshift for $\log(M_{\star}/M_{\odot}) = 11.56$. Bottom: $q_{\mathrm{IR}}$ versus stellar mass at $z = 0.52$.}
    \label{fig:qir_distribution}
\end{figure}
Applying this criterion to the BEAMS sample, we find that the 38 BCGs detected in RACS lie more than 0.43~dex below the mass- and redshift-dependent IRRC at their respective stellar masses and redshifts (see Fig.~\ref{fig:qir_distribution}). This demonstrates that, at the radio luminosities probed by RACS, the detected BCG radio emission cannot be explained by star formation alone and is instead dominated by AGN activity. We therefore classified all RACS-detected BCGs as radio-loud AGNs. We note that BCGs without radio detections cannot be assumed to be AGN-free, as low-luminosity or radiatively inefficient AGNs can fall below the RACS detection limit.

Given the RACS detection threshold ($\sim$5$\sigma \approx 1.3$ mJy), the corresponding radio luminosity limit increases with redshift, from $L_{1.4} \sim 10^{23}$~W~Hz$^{-1}$ at $z \sim 0.3$ to $\sim10^{24}$~W~Hz$^{-1}$ at $z \sim 0.8$. As a result, low-luminosity radio-loud AGNs are progressively missed at higher redshifts, and the measured AGN fraction should be interpreted as a lower limit to the true incidence of radio-loud AGNs. Wide-area studies have shown that radio-AGN fractions derived from flux-limited surveys are systematically underestimated due to sensitivity and identification incompleteness \citep[e.g.][]{hardcastle2025}.

\subsubsection{Excitation classification: HERGs and LERGs}
\label{sec:herg_lerg}

The optical excitation state provides a powerful diagnostic of the dominant accretion mode in radio-loud AGNs. Following \citet{BestHeckman2012}, we distinguished between HERGs and LERGs using the EW of the [O\,\textsc{iii}]~$5007$\,\AA\ emission line, which traces the ionising output of the accretion flow. Sources are classified as HERGs (`quasar-mode') when $\mathrm{EW} - \sigma_{\mathrm{EW}} > 5$\,\AA, while systems with weaker or statistically insignificant [O\,\textsc{iii}] emission are classified as LERGs and associated with radiatively inefficient, `radio-mode' accretion \citep{BestHeckman2012}.

Among the 109 SALT spectra, four BCGs show emission-line features (Sect.~\ref{sec:opticalclass}). Two are detected in RACS and are classified as HERGs. The first, ACT-CL~J0014.9$-$4036 at $z=0.514$, exhibits strong [O\,\textsc{iii}] emission with $\mathrm{EW([O\,\textsc{iii}])}=17.88\pm1.43$\,\AA, well above the HERG threshold. The second is Phoenix~A at $z=0.596$. This is the only emission-line BCG in our sample for which the [O\,\textsc{iii}]~$5007$\,\AA\ line falls outside the SALT spectral coverage. However, extensive multiwavelength observations demonstrate that Phoenix~A hosts a powerful radiatively efficient AGN with a bolometric luminosity of $\sim10^{46}$\,erg\,s$^{-1}$ and an Eddington ratio of $\sim0.1$--0.3 \citep{McDonald2012,McDonald2015,HlavacekLarrondo2015,Russell2017}, firmly placing it in the HERG regime.

The remaining two emission-line BCGs show substantially weaker [O\,\textsc{iii}] emission ($2 \lesssim \mathrm{EW([O\,\textsc{iii}])} \lesssim 5$\,\AA) than the HERGs. ACT-CL~J0159.2+0030 at $z=0.381$ is detected in RACS, but its [O\,\textsc{iii}] emission lies well below the HERG threshold and it is therefore classified as a LERG. In contrast, ACT-CL~J0234.7$-$5831 at $z=0.415$ is undetected in RACS and likewise exhibits only weak [O\,\textsc{iii}] emission. Given the absence of radio emission and the ambiguity of the ionisation mechanism at such low EWs, we do not classify this source as an AGN and excluded it from the AGN sample considered in this work. The final radio-loud AGN sample therefore comprises 2 HERGs and 36 LERGs.

\subsection{Mid-infrared classification}
\label{sec:mirclass}

Mid-infrared colours provide a widely used diagnostic of radiatively efficient AGN activity, as they trace thermal emission from warm and hot dust heated by accretion processes \citep[e.g.][]{Stern2012,Assef2018}. In particular, the WISE $W1$, $W2$, and $W3$ bands are sensitive to emission from AGN-heated dust in a circumnuclear torus, allowing obscured and unobscured radiatively efficient AGNs to be identified even when optical signatures are weak or absent. We therefore used the $W1-W2$ versus $W2-W3$ colour-colour plane, constructed from AllWISE photometry (Sect.~\ref{sec:mirdata}), as an independent probe of radiative AGN activity in the BEAMS BCG sample. The $k$-corrected WISE colours used in this analysis were derived from a dedicated catalogue provided by the late Prof.~T.~H.~Jarrett\footnote{Private communication.}.

\begin{figure}
    \centering
    \includegraphics[width=0.92\columnwidth]{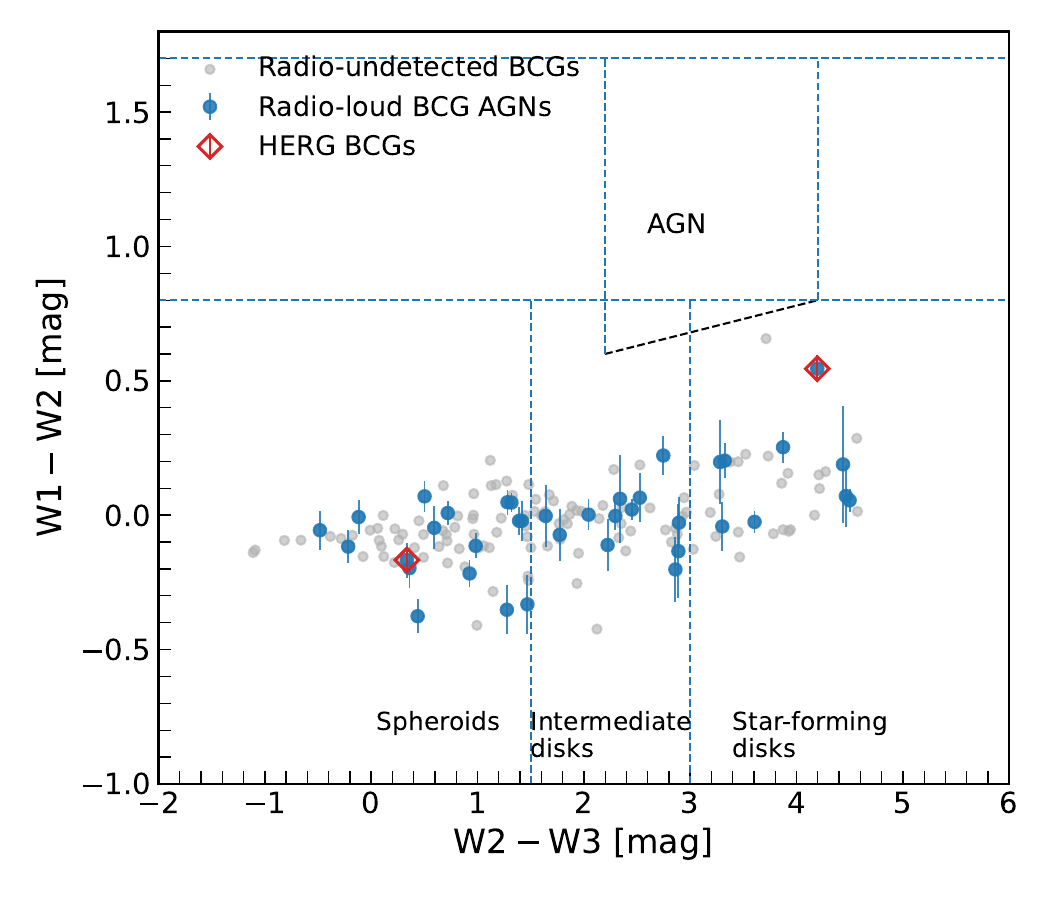}
    \caption{WISE MIR colour--colour diagram of the BEAMS BCGs. Grey points show BCGs with WISE detections, blue circles the RACS-detected subset, and red diamonds the optically identified HERGs. Dashed lines indicate empirical MIR colour regions defined for the general galaxy population (spheroid, intermediate, star-forming, and AGN-dominated; \citealt{Yao2020}). Error bars indicate the 1$\sigma$ uncertainties on $W1-W2$.}
    \label{fig:wise_color}
\end{figure}

Studies of WISE colour-colour space have shown that massive galaxies hosting weak or obscured AGNs often occupy an intermediate region between the star-forming sequence and the canonical AGN wedge \citep{Jarrett2011}. This `mWarm' population, identified by \citet{Yao2020}, may include systems with modest AGN-heated dust emission, although star formation and evolved stellar populations can also contribute to the observed MIR colours. Figure~\ref{fig:wise_color} shows that most BEAMS BCGs have WISE colours within the empirical spheroid and intermediate-disk (mWarm) regions, with a smaller subset extending towards the star-forming locus. These regions describe the MIR colours of the general galaxy population and should not be interpreted as a classification scheme for radio AGN hosts. Indeed, radio-loud BCG AGNs span a broad range of WISE colours, consistent with \citet{Gurkan2014}, who showed that radio AGNs cannot be reliably classified using WISE colour-colour diagrams alone. No BCG in our sample satisfies the canonical MIR AGN criterion of $W1-W2>0.8$. Phoenix A, however, is located close to the empirical AGN-dominated region and well away from the quiescent spheroid sequence, consistent with its independent HERG classification.

The absence of MIR-selected AGNs despite the presence of optically and radio-confirmed systems further illustrates the incompleteness of MIR colour selection for AGNs hosted by massive cluster centrals. This is consistent with recent work showing that WISE colour selection is generally unreliable for identifying radio-loud AGNs in wide-area surveys \citep[e.g.][]{hardcastle2025}, and suggests that this limitation is particularly relevant for BCGs. In these systems, the combination of extreme stellar mass and predominantly low-Eddington accretion likely leads to intrinsically weak and/or strongly diluted hot-dust emission from the torus \citep[e.g.][]{Donley2012,Mendez2013,Hardcastle2016}.

\section{Results}
\label{sec:results}

\begin{figure}
    \centering
    \includegraphics[width=0.83\columnwidth]{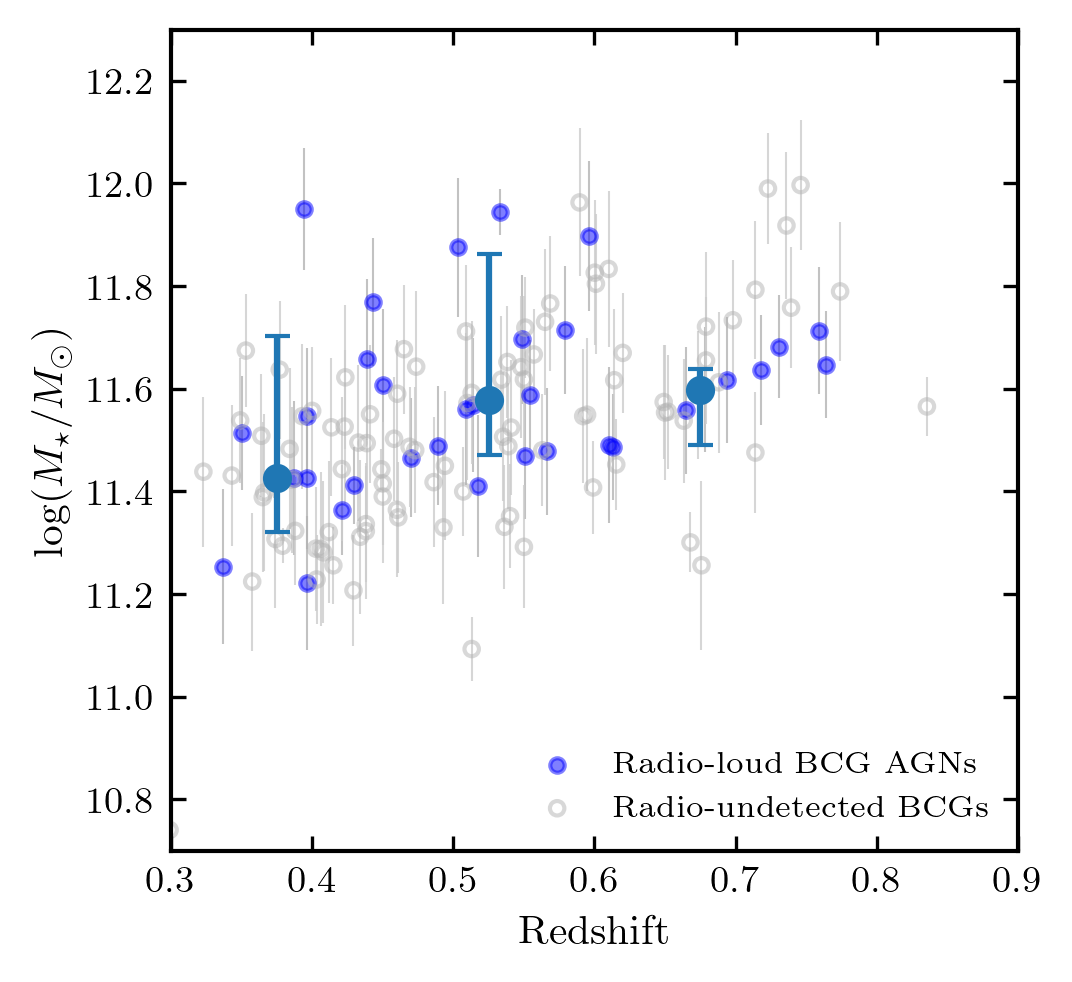} 
    \includegraphics[width=0.83\columnwidth]{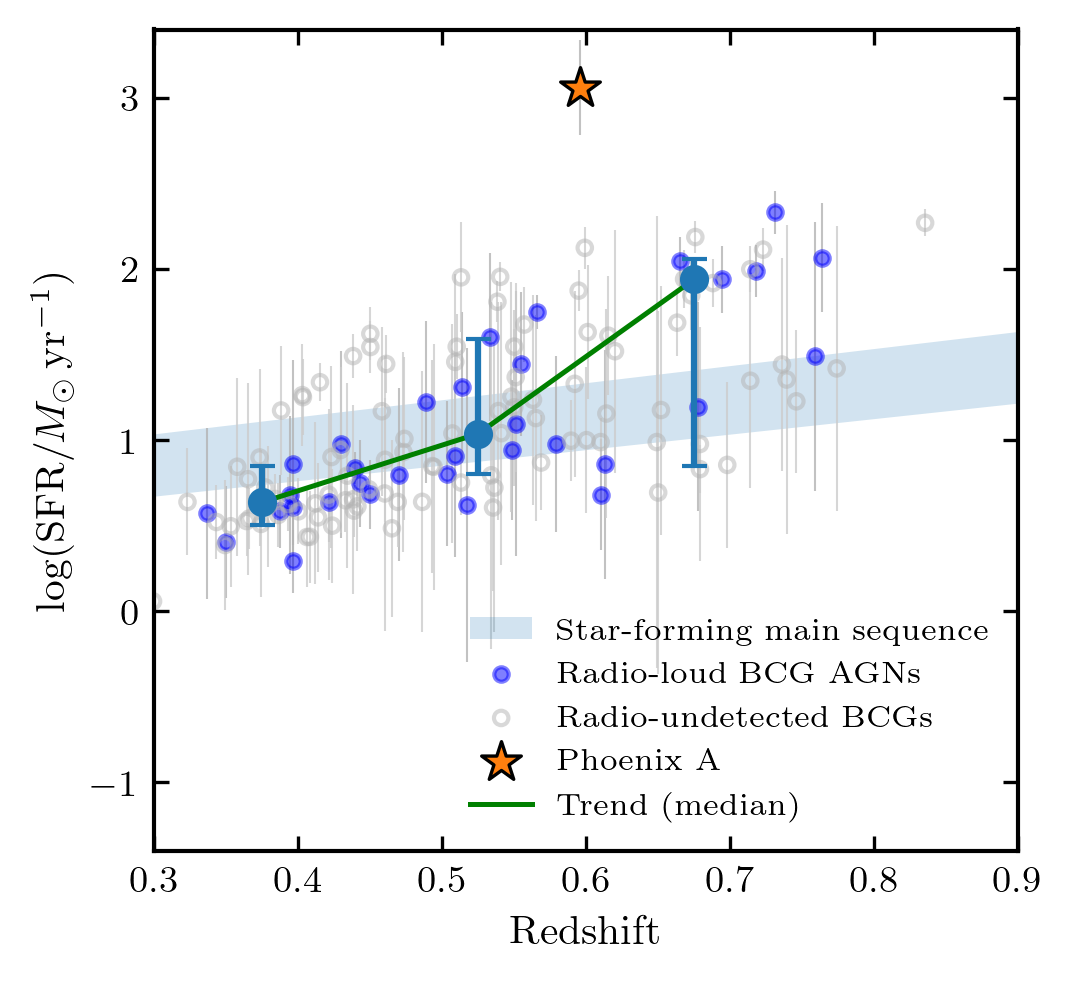}
    \caption{Host-galaxy properties of the BEAMS BCGs versus redshift.
    \textit{Top}: Stellar mass.
    \textit{Bottom}: SFR. Blue points show the 34 radio-loud AGN hosts and grey circles the radio-undetected BCGs.
    Individual points include $1\sigma$ uncertainties from the \textsc{X-CIGALE} fits. Filled symbols with vertical error bars show the median values of the radio-loud subsample in redshift bins, with the 16th--84th percentile range.
    The shaded region in the bottom panel indicates the star-forming main sequence over the stellar mass range of the sample ($11.2 \lesssim \log(M_\star/M_\odot) \lesssim 11.9$; \citealt{Speagle2014}) at $0.3 < z < 0.8$.}
    \label{fig:bcg_mz_sfrz}
\end{figure}

\subsection{Host-galaxy properties of radio-loud BCG AGNs}
\label{sec:host_properties}

Of the full sample of 171 BCGs, 38 systems are classified as radio-loud AGNs (Sect.~\ref{sec:radioclass}). Of these 38, 34 have robust SED-fitting results and are therefore used in the analysis of SED-derived parameters, such as stellar mass and SFR. We focused primarily on the host-galaxy properties of these radio-detected systems, examining whether galaxies that host radio-mode activity exhibit systematic trends in stellar mass or SFR across $0.3 < z < 0.8$. Where relevant, we compared these trends with those observed in BCGs without radio detections to assess whether they are specific to AGN hosts or reflect the general evolution of the BCG population.

\subsubsection{Stellar mass}

The stellar masses of the radio-loud BCG AGNs are shown as a function of redshift in Fig.~\ref{fig:bcg_mz_sfrz}. Throughout the entire redshift interval, the AGN-hosting BCGs remain uniformly massive, spanning $11.22 \leq \log(M_\star/M_\odot) \leq 11.95$, with a median value of $\log(M_\star/M_\odot) = 11.56$. The 16th--84th percentile range is $11.43$--$11.71$, indicating a relatively narrow distribution.

A Spearman rank test\footnote{The Spearman rank correlation coefficient ($\rho$) measures the strength of a monotonic relationship between two variables based on their ranked values. It is equivalent to the Pearson correlation applied to the rank-transformed data. The associated $p$-value gives the probability of obtaining such a correlation under the null hypothesis of no association.} reveals a weak-to-moderate positive correlation between stellar mass and redshift, with $\rho = 0.40$ and $p = 0.020$. The total variation amounts to $\lesssim 0.3$ dex over $0.3 < z < 0.8$. For comparison, BCGs without radio detections exhibit a stronger correlation between stellar mass and redshift ($\rho = 0.56$, $p \simeq 1.6 \times 10^{-9}$), indicating more pronounced apparent evolution in the full population. However, this trend is unlikely to reflect intrinsic stellar-mass growth. Instead, it is most plausibly driven by selection effects, as SZ-selected cluster samples preferentially include more massive haloes at higher redshifts, which in turn host more massive BCGs.

The relatively weak mass evolution observed for radio-detected systems therefore suggests that these AGNs preferentially reside in already assembled, extremely massive galaxies. This is consistent with scenarios in which much of the stellar mass in BCGs was established by $z \sim 0.8$, with subsequent evolution dominated by minor mergers and satellite accretion rather than sustained in situ star formation \citep[e.g.][]{Lidman2012,Lin2013,Burke2015}. We note that BCGs without radio detections can still host low-luminosity AGNs below the RACS detection limit.

\subsubsection{Star formation rates}

The SFRs of the radio-loud BCG AGNs span $0.29 \leq \log(\mathrm{SFR}/M_\odot\,\mathrm{yr}^{-1}) \leq 3.06$, with a median value of $\log(\mathrm{SFR}/M_\odot\,\mathrm{yr}^{-1}) = 0.92$ and a 16th--84th percentile range of $0.63$--$1.89$. As shown in Fig.~\ref{fig:bcg_mz_sfrz}, the BCG population lies both below and around the evolving star-forming main sequence at comparable redshifts, with a substantial fraction of systems consistent with main-sequence star formation. The shaded region in Fig.~\ref{fig:bcg_mz_sfrz} represents the expected main-sequence relation over the stellar mass range of the BCG sample ($11.2 \lesssim \log(M_\star/M_\odot) \lesssim 11.9$), based on the redshift-dependent parametrisation of \citet{Speagle2014}. Individual offsets from the main sequence are evaluated relative to each galaxy's stellar mass, so the shaded region illustrates the expected range of main-sequence loci across the sample rather than a strict classification boundary.

In contrast to the relatively weak stellar-mass evolution, we find a strong correlation between SFR and redshift. A Spearman rank test yields $\rho = 0.80$ with $p = 1.2 \times 10^{-8}$, indicating that SFR increases systematically towards higher redshifts. The typical SFR rises by nearly an order of magnitude between $z \sim 0.3$ and $z \sim 0.8$. This trend is not driven solely by the extreme Phoenix~A system: excluding Phoenix~A yields nearly identical results ($\rho = 0.81$, $p = 1.1 \times 10^{-8}$), demonstrating that the observed evolution reflects the broader radio-loud BCG population.
BCGs without radio detections also show a strong SFR--redshift correlation ($\rho = 0.67$, $p \simeq 3.7 \times 10^{-14}$), indicating that this behaviour reflects the general evolution of the BCG population. However, the correlation is somewhat stronger for radio-detected systems, suggesting that BCGs that host radio-loud AGNs more closely track the increase in gas availability with redshift, although the difference is modest. A similar trend is observed for the specific SFR, which correlates strongly with redshift for radio-detected BCGs ($\rho = 0.72$, $p = 1.3 \times 10^{-6}$), compared with a weaker but still significant correlation for radio-undetected systems ($\rho = 0.43$, $p = 7.8 \times 10^{-6}$). This suggests that the enhanced evolution is not driven solely by stellar-mass differences, but also reflects increasing star-formation activity relative to galaxy mass towards higher redshifts. We note that radio-undetected BCGs can still host low-luminosity AGNs below the RACS detection limit.

This behaviour mirrors the broader cosmic decline in cold-gas fractions and star-formation activity since $z\sim1$ \citep[e.g.][]{MadauDickinson2014}. Our SFR estimates are derived from UV-to-MIR photometry using \textsc{X-CIGALE} and are not directly constrained by FIR observations. As discussed in Sect.~\ref{sec:mirdata}, our FIR sensitivity test shows that including SPIRE data can substantially lower the inferred SFRs; for the FIR-covered subset, the median offset from the star-forming main sequence, defined as $\Delta_{\rm MS} \equiv \log({\rm SFR}_{\rm BCG})-\log[{\rm SFR}_{\rm MS}(M_\star,z)]$, changes from $-0.05$ dex to $-1.57$ dex. However, the FIR-inclusive SFR uncertainties are generally large, and none of the FIR-inclusive fits satisfies the $\chi^2_{\rm r}<3$ criterion adopted for the main analysis. We therefore interpret the absolute SFR normalisation and individual positions relative to the main sequence with caution.
At higher redshifts, the DES optical bands probe progressively bluer rest-frame wavelengths, increasing sensitivity to UV emission that can be interpreted by \textsc{X-CIGALE} as enhanced recent star formation if dust attenuation is not tightly constrained \citep[e.g.][]{Ciesla2015,Boquien2019}. In addition, AGN-heated dust emission in the WISE $W3$ and $W4$ bands becomes increasingly difficult to separate from star-forming dust emission as prominent polycyclic aromatic hydrocarbon features shift through the observed MIR bands \citep[e.g.][]{Donley2012,Kirkpatrick2015,Masoura2018}. These effects may contribute to systematic uncertainties in the SFRs of some passive systems with weak MIR detections. Nevertheless, the strong redshift dependence persists even after excluding Phoenix~A, indicating that the observed trend is not driven solely by a small number of outliers. Our results therefore indicate that the radio-loud BCG population spans a broad range of star-formation activity rather than being confined to uniformly quenched systems.

\subsubsection{Implications for AGN fuelling}

The combined trends in stellar mass and SFR indicate that radio-mode AGN activity in BCGs is not associated with substantial ongoing stellar-mass growth. AGN hosts remain among the most massive galaxies across the full redshift range considered, exhibiting only weak stellar-mass evolution. By contrast, the strong increase in SFR towards higher redshifts suggests evolving cold-gas properties and/or star-formation efficiency.

The stronger evolution in SFR relative to stellar mass supports a scenario in which black hole fuelling responds to changing gas availability without driving significant new stellar-mass assembly. However, higher SFRs at earlier epochs do not necessarily imply proportionally larger cold-gas reservoirs, as galaxies at high redshifts are also known to exhibit enhanced star-formation efficiencies \citep[e.g.][]{Tacconi2020}. Although systematic uncertainties in MIR-based SFR estimates may affect the absolute normalisation of the observed evolution (Sect.~\ref{sec:host_properties}), the persistence of the trend in both radio-detected and radio-undetected BCGs suggests that it reflects genuine cosmic evolution in the baryonic properties of massive cluster galaxies.

Within this framework, radio-mode AGN activity represents a persistent maintenance mechanism operating in already massive systems \citep[e.g.][]{McNamaraNulsen2012}, while the gradual increase in SFR towards higher redshifts is consistent with the broader evolution of gas fractions and star-formation activity in massive haloes. The broad distribution of SFRs, located both below and around the star-forming main sequence, further indicates that radio-mode AGNs can be hosted by galaxies with a wide range of star-formation activity rather than being confined to strongly quenched systems.

\subsection{Radio properties of BCG AGNs}
\label{sec:results_radio}

The 38 radio-loud BCG AGNs are overwhelmingly dominated by LERGs, with only two systems classified as HERGs (Sect.~\ref{sec:herg_lerg}). We therefore focused on the properties of this predominantly radio-mode population and examined how their radio luminosities relate to host-galaxy star formation.

\begin{figure}
    \centering
    \includegraphics[width=0.85\columnwidth]{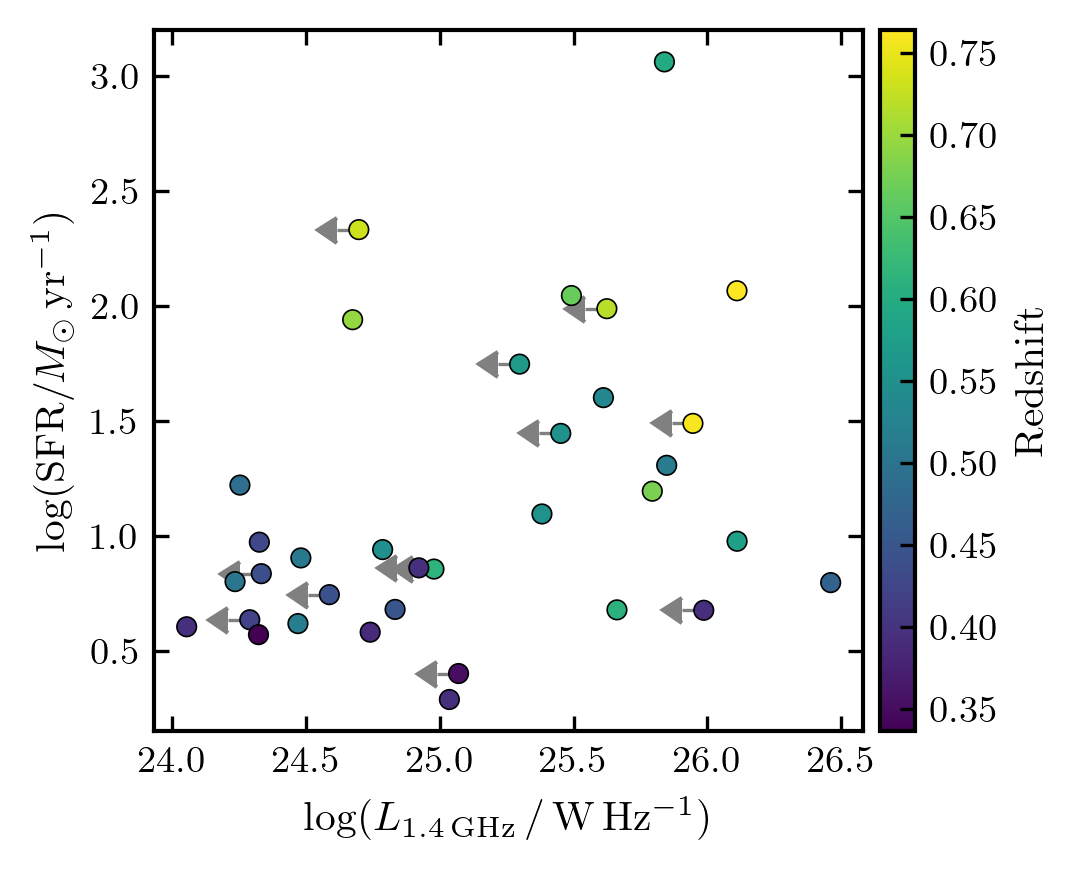}
    \caption{Rest-frame 1.4\,GHz radio luminosity versus SFR for the 38 radio-loud AGN BCGs. Points are colour-coded by redshift. A moderate positive trend is present, although both quantities increase with redshift, as indicated by the colour gradient. Grey arrows denote upper limits for systems where the radio emission is blended with nearby sources or offset from the BCG optical position, making the radio luminosities uncertain.}
    \label{fig:Lradio_SFR_z}
\end{figure}

We find a moderate positive correlation between radio luminosity and SFR among the radio-loud BCG AGNs (Spearman $\rho = 0.38$, $p = 0.026$). However, as shown in Fig.~\ref{fig:Lradio_SFR_z}, both quantities increase systematically with redshift across the BEAMS sample. Radio luminosity exhibits a significant redshift dependence ($\rho = 0.45$, $p = 0.007$), while SFR shows even stronger evolution ($\rho = 0.80$, $p \ll 10^{-6}$). Excluding the extreme Phoenix~A system yields a similar, though slightly weaker, relation between radio luminosity and SFR ($\rho = 0.35$, $p = 0.045$), indicating that the observed trend is not dominated by a single outlier. The apparent $L_{1.4}$--SFR relation is therefore likely driven primarily by the shared redshift evolution of the two quantities, rather than indicating an intrinsic coupling between jet power and ongoing star formation.
We therefore find no compelling evidence that radio jet power scales directly with the cold-gas reservoir, as traced indirectly by the global SFR. Instead, both radio luminosity and star-formation activity appear to respond to the broader evolution of gas supply and/or star-formation efficiency in cluster cores, consistent with a scenario in which radio-mode accretion is linked to the thermodynamic state of the ICM \citep[e.g.][]{McNamara2007,Fabian2012,Voit2015}.

\subsubsection{Comparison with MIGHTEE-COSMOS radio-AGN population}

\begin{figure}
    \centering
    \includegraphics[width=0.94\columnwidth]{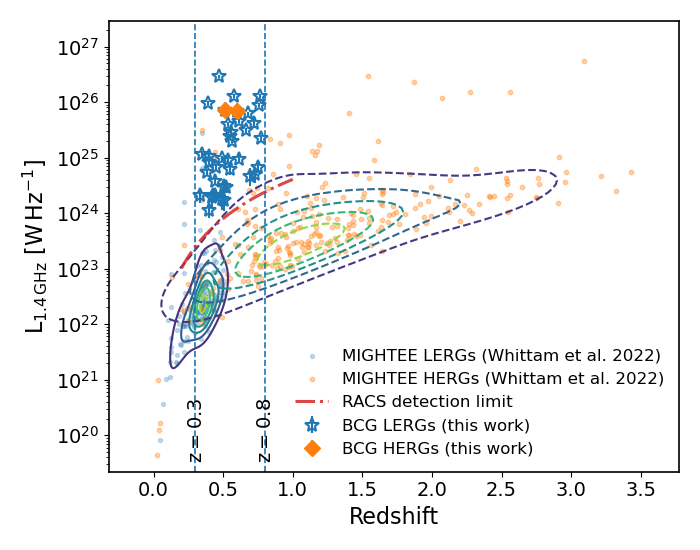}
    \caption{
    Radio power at 1.4\,GHz versus redshift for the BEAMS BCGs compared with the MIGHTEE-COSMOS radio-loud AGN sample \citep{Whittam2022}. Blue and orange points show MIGHTEE-COSMOS LERGs and HERGs, respectively; solid and dashed contours indicate their two-dimensional density distributions in the redshift--luminosity plane. Stars denote the BCG LERGs and diamonds the two BCG HERGs.
    }
    \label{fig:mightee-compare}
\end{figure}

\begin{figure}
    \centering
    \includegraphics[width=0.79\columnwidth]{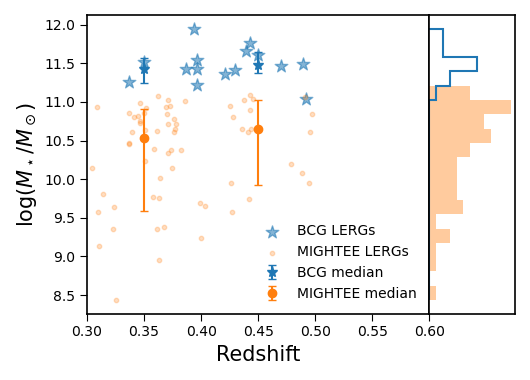}
    \includegraphics[width=0.79\columnwidth]{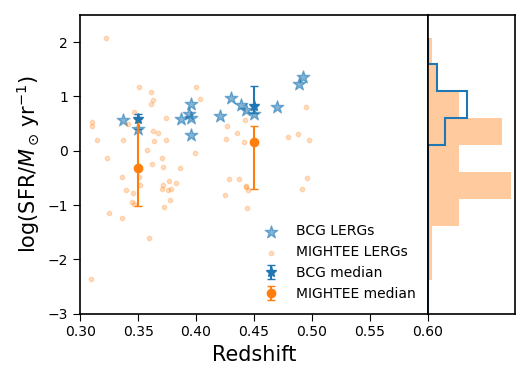}
    \caption{Comparison of the stellar masses and SFRs of BCG LERGs with field LERGs from the MIGHTEE-COSMOS survey over $0.3 < z < 0.5$. \textit{Top}: Stellar mass versus redshift. \textit{Bottom}: SFR versus redshift. Light blue points show the MIGHTEE-COSMOS LERGs and orange stars the BCG LERGs. Filled symbols with error bars show the median values of the BCG sample in redshift bins, with the 16th--84th percentile range. Histograms on the right show the corresponding distributions.
    }
    \label{fig:mightee_lerg_m_sfr}
\end{figure}

To determine whether radio-loud AGNs hosted by BCGs are distinct from the general radio-AGN population at similar cosmic epochs, we compared the BEAMS radio-loud BCG AGNs with HERGs and LERGs drawn from the MIGHTEE-COSMOS field survey catalogue \citep[][hereafter MIGHTEE-COSMOS]{Whittam2022}. We first examined their position in radio luminosity-redshift space (Fig.~\ref{fig:mightee-compare}) before considering host-galaxy stellar masses and SFRs (Fig.~\ref{fig:mightee_lerg_m_sfr}). To enable a like-for-like statistical comparison of host properties, we restricted the analysis to LERGs and to the common redshift interval $0.3 < z < 0.5$ in MIGHTEE-COSMOS, where the LERG sample is well represented in the catalogue. At higher redshifts, the fraction of sources with robust LERG classifications decreases, leading to increased incompleteness.

\paragraph{Radio power.}

Figure~\ref{fig:mightee-compare} shows the rest-frame 1.4\,GHz radio power as a function of redshift for the BEAMS BCGs and the MIGHTEE-COSMOS HERGs and LERGs. The BCG radio sources occupy the upper envelope of the luminosity distribution at fixed redshift, with most systems lying in the range $\log(L_{1.4}/{\rm W\,Hz^{-1}})\sim24$--26. However, this is largely driven by the RACS flux limit, which restricts the sample to moderate- and high-power radio sources. At the median redshift of the BEAMS sample ($z\sim0.5$), the $5\sigma$ RACS detection threshold corresponds to $\log(L_{1.4}/{\rm W\,Hz^{-1}})\approx23.3$, rendering lower-luminosity HERGs and LERGs undetectable.

Restricting the comparison to MIGHTEE-COSMOS sources above the same effective luminosity limit and within $0.3 < z < 0.8$, the BCG LERGs are statistically consistent with the field LERG population. A KS test yields $p = 0.16$, indicating no significant difference between the radio luminosity distributions. Although the BCG LERGs exhibit a moderately higher median luminosity, by $\sim0.4$ dex, they remain consistent with the upper end of the field population rather than forming a distinct high-luminosity population. The two BCG HERGs likewise lie towards the luminous end of the field HERG distribution but remain consistent with the upper envelope given the small sample size.
Overall, we find no compelling evidence that BCG radio-loud AGNs are intrinsically over-luminous in radio power at a fixed epoch. Instead, the observed luminosity distribution is strongly shaped by the flux-limited nature of the BEAMS sample detected in RACS.

\paragraph{Stellar mass.}

In contrast, the host-galaxy stellar masses show a clear systematic offset (top panel of Fig.~\ref{fig:mightee_lerg_m_sfr}). Over $0.3 < z < 0.5$, MIGHTEE-COSMOS LERGs have a median $\log(M_\star/M_\odot)=10.62$ (16th--84th percentile: 9.66--10.93), whereas BCG LERGs have a median $\log(M_\star/M_\odot)=11.47$ (11.28--11.65), corresponding to an offset of $\sim0.85$ dex. A KS test yields $D=0.95$ with $p=2.5\times10^{-13}$, demonstrating that the stellar-mass distributions are statistically distinct.

The MIGHTEE-COSMOS host-galaxy properties were derived using \textsc{LePhare} \citep{Arnouts2011}, while the BEAMS values were obtained with \textsc{X-CIGALE}. Comparisons between SED-fitting codes indicate that stellar masses are generally robust to within $\sim0.1$ dex and rarely differ by more than $\sim0.2$--0.3 dex for comparable assumptions \citep[e.g.][]{Santini2015,Pacifici2023}. The observed $\sim0.85$ dex offset therefore substantially exceeds plausible methodological systematics and indicates that BCG LERGs are intrinsically more massive than field LERGs at similar redshifts. This is consistent with previous studies showing that radio-loud AGNs preferentially inhabit massive galaxies \citep[e.g.][]{hardcastle2025}, with BCGs occupying the high-mass end of the LERG population.

\paragraph{Star-formation rates.}

The SFR distributions also differ significantly (bottom panel of Fig.~\ref{fig:mightee_lerg_m_sfr}). MIGHTEE-COSMOS LERGs have a median $\log(\mathrm{SFR}/M_\odot\,\mathrm{yr}^{-1})=-0.31$ (16th--84th percentile: $-0.94$ to 0.56), whereas BCG LERGs have a higher median value of $0.68$ (0.58--0.95). A KS test yields $D=0.71$ with $p=1.4\times10^{-6}$.
However, SFR estimates are more sensitive to SED-model assumptions than stellar masses. In particular, \textsc{LePhare} does not implement an energy-balance treatment of dust emission, whereas \textsc{X-CIGALE} incorporates MIR constraints. Inter-comparisons of SED-fitting methods show that SFRs can differ systematically by $\sim0.3$ dex and, in some cases, by up to $\sim0.5$ dex \citep{Pacifici2023}. Because the BEAMS SFRs are derived from optical+MIR data without FIR constraints, while the MIGHTEE-COSMOS SFRs use broader multiwavelength coverage, part of the observed offset may therefore reflect methodological differences in addition to any genuine environmental effect.

Taken together, these results indicate that BCG radio-loud AGNs are not distinguished by their radio-power distribution, but instead represent the high-mass end of the general LERG population. Their radio powers are broadly consistent with those of field LERGs once selection effects are taken into account, while their host galaxies are systematically more massive and exhibit moderately higher inferred SFRs.

\subsection{Accretion rates}
\label{sec:results_accretion}

\begin{figure}
    \centering
    \includegraphics[width=0.8\columnwidth]{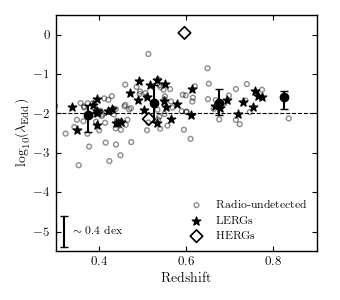}
    \caption{Eddington-scaled AGN accretion rate versus redshift for the BEAMS BCG sample.
    Grey open circles show radio-undetected BCGs, black stars the radio-detected LERGs, and open diamonds the HERGs.
    The dashed horizontal line marks $\lambda_{\mathrm{Edd}} = 10^{-2}$, the conventional boundary between radiatively efficient and inefficient accretion.
    Filled circles with vertical error bars show the median $\log_{10}(\lambda_{\mathrm{Edd}})$ of the full BCG sample in redshift bins, with the 16th--84th percentile range.
    }
    \label{fig:acc_rate_z}
\end{figure}

\begin{figure}
    \centering
    \includegraphics[width=0.78\columnwidth]{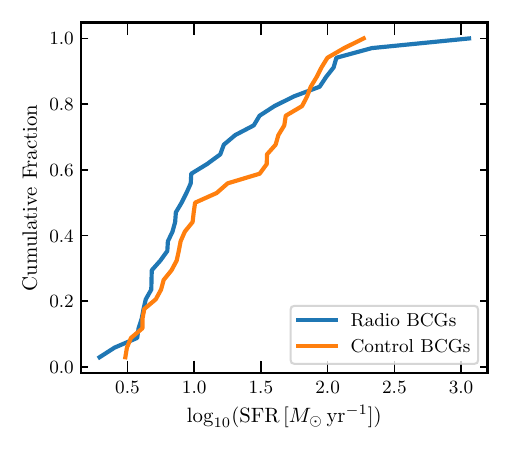}
    \includegraphics[width=0.78\columnwidth]{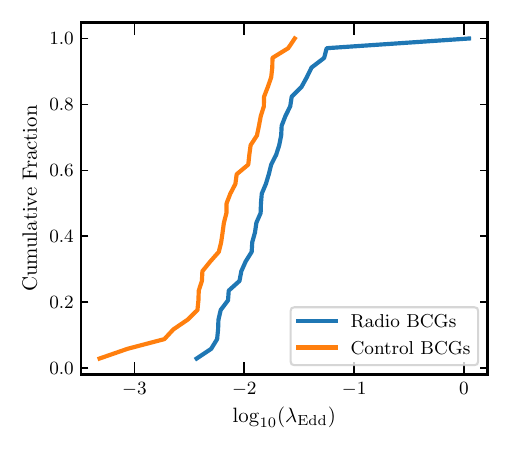}
    \includegraphics[width=0.78\columnwidth]{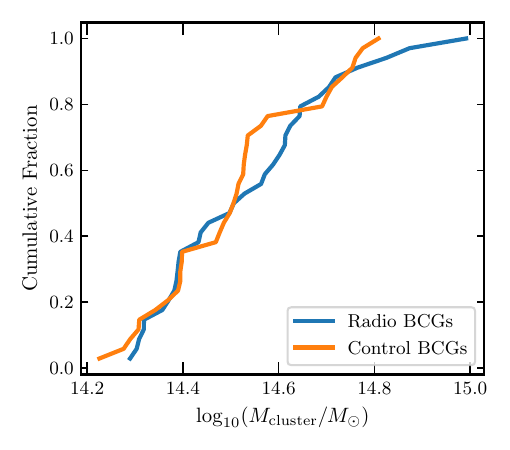}
    \caption{Cumulative distribution functions of the radio-detected BCGs (red) and a stellar-mass- and redshift-matched control sample (blue).
    \textit{Top}: SFR.
    \textit{Middle}: Eddington-scaled accretion rate.
    \textit{Bottom}: Host cluster mass.
    Only the Eddington-ratio distributions differ significantly (KS $p=0.028$); the SFR and cluster-mass distributions are statistically indistinguishable.
    }
    \label{fig:cdf_radio_control}
\end{figure}

The Eddington-scaled accretion rate provides a physically motivated measure of the instantaneous growth efficiency of supermassive black holes, linking observable AGN output to the underlying accretion mode. Radiatively efficient, quasar-mode AGNs typically accrete at $\lambda_{\mathrm{Edd}} \gtrsim 10^{-2}$, while radio-mode, jet-dominated systems are characterised by lower Eddington ratios, $\lambda_{\mathrm{Edd}} \lesssim 10^{-2}$, consistent with radiatively inefficient accretion flows \citep[e.g.][]{BestHeckman2012,HeckmanBest2014,Sabater2019,Whittam2022}. In BCGs, where AGN feedback operates predominantly in a maintenance mode regulated by the ICM, low Eddington ratios are therefore expected \citep[e.g.][]{Hardcastle2007}.

Bolometric luminosities ($L_{\mathrm{bol}}$) were taken from the Bayesian AGN accretion power (\texttt{bayes.agn.accretion\_power}) returned by \textsc{X-CIGALE}, and black hole masses were estimated from stellar masses using $M_{\mathrm{BH}} = 0.0014\,M_{\star}$ \citep{Haering2004}. The Eddington ratio was then computed as $\lambda_{\mathrm{Edd}} = L_{\mathrm{bol}}/L_{\mathrm{Edd}}$, where $L_{\mathrm{Edd}} = 1.3\times10^{31}(M_{\mathrm{BH}}/M_{\odot})$\,W. Systematic uncertainties are of order $\sim0.4$ dex, dominated by the scatter in the $M_{\mathrm{BH}}$--$M_\star$ relation and uncertainties in $L_{\mathrm{bol}}$. Black hole masses derived from the $M_{\mathrm{BH}}$--$\sigma_\star$ relation \citep{Tremaine2002} are consistent within these uncertainties (Appendix~\ref{appendix:mbh_comparison}).

The Eddington ratio is defined here using only the radiative component of the AGN output. While this provides a useful proxy for radiatively efficient AGNs, it should be interpreted with caution for radiatively inefficient systems (LERGs). In such sources, the nuclear optical and X-ray emission is thought to arise predominantly from jet-related emission, while the radiative emission from the accretion flow itself is expected to be intrinsically weak \citep[e.g.][]{Chiaberge1999,HardcastleWorrall2000}. Consequently, the AGN bolometric luminosities inferred from SED fitting should be regarded as upper limits for radiatively inefficient systems rather than direct measurements of the accretion luminosity. The corresponding Eddington ratios should therefore likewise be interpreted as upper limits for LERGs. Although the inclusion of jet mechanical power would provide a more complete estimate of the total AGN power, doing so would require uncertain scaling relations and would restrict the analysis to the radio-detected subsample, precluding a uniform comparison across the full sample.

Figure~\ref{fig:acc_rate_z} shows $\log_{10}(\lambda_{\mathrm{Edd}})$ as a function of redshift for the 133 BCGs. The distribution spans predominantly low to intermediate values, with most systems occupying $-3 \lesssim \log_{10}(\lambda_{\mathrm{Edd}}) \lesssim -1$, consistent with radiatively inefficient accretion. While 64\% of sources lie formally above $\lambda_{\mathrm{Edd}} = 10^{-2}$, this should not be over-interpreted. For the LERG population, the SED-derived bolometric luminosities, and hence the inferred Eddington ratios, should be regarded as upper limits rather than direct measurements of the true radiative accretion luminosity. The apparent fraction of sources above the nominal radiatively efficient threshold is therefore likely to be overestimated. Even these systems remain well below the Eddington ratios typical of luminous quasars, which generally accrete at $\lambda_{\mathrm{Edd}} \sim 0.1$--1 \citep[e.g.][]{Kollmeier2006,Shen2011}.
Radio-detected and radio-undetected BCGs span similar ranges in $\lambda_{\mathrm{Edd}}$, with substantial overlap at fixed redshift. The radio-undetected population extends to lower values, but this is likely driven by selection effects: the flux-limited nature of RACS biases against low-luminosity, low-accretion-rate AGNs at higher redshifts. The two HERGs occupy the upper end of the distribution, although one lies below $\log_{10}(\lambda_{\mathrm{Edd}}) = -2$; given the uncertainties and known overlap between accretion modes, this difference is not significant. We find a moderate increase in $\lambda_{\mathrm{Edd}}$ with redshift (Spearman $\rho = 0.38$, $p = 4.9 \times 10^{-6}$), which persists when considering only radio-loud AGNs ($\rho = 0.40$, $p = 0.018$). This evolution is consistent with increased gas availability at earlier epochs.

Overall, the accretion properties of radio-detected BCG AGNs are similar to those of field radio-loud AGNs \citep[see Fig.~12 in][]{Whittam2022}, indicating that they represent the high-mass end of a common accretion population. We note that radio-undetected BCGs can still host low-level or quiescent AGNs below the detection threshold and are therefore not necessarily inactive. These results support a scenario in which AGN activity in BCGs is governed by gas supply regulated by the ICM, rather than by extreme star-formation or quenching states.

\subsubsection{Radio power versus accretion efficiency}

Having established the redshift evolution of $\lambda_{\mathrm{Edd}}$, we examined whether radio jet power scales with radiative accretion efficiency. Comparing rest-frame 1.4\,GHz radio luminosity with $\lambda_{\mathrm{Edd}}$ for the radio-loud BCG AGNs, we find no statistically significant correlation (Spearman $\rho = 0.09$, $p = 0.63$). Radio power therefore does not show a clear dependence on the instantaneous radiative accretion efficiency. Such behaviour is characteristic of kinetic-mode AGNs, in which mechanically dominated outflows persist across a broad range of low to intermediate Eddington ratios \citep[e.g.][]{Hardcastle2007,BestHeckman2012,HeckmanBest2014,Sabater2019,Whittam2022}. Studies of radio galaxies in both field and cluster environments find that jet power exhibits substantial scatter at fixed radiative luminosity, consistent with radiatively inefficient accretion flows \citep{Balmaverde2008,Russell2013}. These results indicate that radio power in BCGs is not governed solely by radiative accretion rate, but is instead likely regulated by the availability and thermodynamic state of gas in the ICM, which influences the fuelling of the central engine.

\subsection{Radio-detected versus control BCGs}

To assess whether radio activity is associated with systematic differences in host properties, we constructed a stellar-mass- and redshift-matched control sample of 34 radio-undetected BCGs drawn from the parent sample. We required robust SED fits and low AGN fractional contributions ($f_{\mathrm{AGN}} < 0.3$), where $f_{\mathrm{AGN}}$ denotes the fraction of infrared luminosity attributed to AGN emission in the \textsc{X-CIGALE} fits. Matching was performed using a nearest-neighbour approach in $M_\star$ and redshift, yielding excellent agreement, with mean offsets of $\langle \Delta z \rangle = 0.005$ and $\langle \Delta \log M_* \rangle = 0.02$~dex, and scatters of $\sigma(\Delta z)=0.07$ and $\sigma(\Delta \log M_*)=0.06$~dex. Figure~\ref{fig:cdf_radio_control} shows the cumulative distribution functions of the SFR, Eddington-scaled accretion rate, and host cluster mass for the radio-loud and control samples.

The SFR distributions (Fig.~\ref{fig:cdf_radio_control}a) exhibit substantial overlap. The median values are $\log_{10}(\mathrm{SFR}/M_\odot\,\mathrm{yr}^{-1}) = 0.92$ and $1.09$ for the radio-loud and control samples, respectively, with bootstrap ranges of $0.92^{+0.11}_{-0.08}$ and $1.09^{+0.40}_{-0.13}$ dex (16th--84th percentiles). A two-sample KS test ($D = 0.21$, $p = 0.47$) indicates no statistically significant difference, implying that radio-loud BCGs do not exhibit enhanced or suppressed star formation relative to matched radio-undetected systems.

In contrast, the Eddington-scaled accretion rates (Fig.~\ref{fig:cdf_radio_control}b) show a systematic offset. Both samples occupy the low to intermediate Eddington regime, with the radio-loud BCG AGNs shifted towards higher values. The median $\log_{10}(\lambda_{\mathrm{Edd}})$ is $-1.84$ for the radio-loud sample and $-2.14$ for the control sample, with bootstrap ranges of $-1.84^{+0.08}_{-0.05}$ and $-2.14^{+0.13}_{-0.04}$ dex, respectively. A KS test ($D = 0.35$, $p = 0.028$) indicates a statistically significant difference. Consistently, $71\%$ of radio-loud BCG AGNs have $\log_{10}(\lambda_{\mathrm{Edd}}) > -2$, compared with $41\%$ of the control sample. Excluding the most extreme system, Phoenix~A, does not alter this result.

The cluster mass distributions (Fig.~\ref{fig:cdf_radio_control}c) are statistically indistinguishable, with median $\log_{10}(M_{\mathrm{cluster}}/M_\odot) = 14.52$ and $14.51$ dex for the radio and control samples, respectively. A KS test ($D = 0.21$, $p = 0.47$) confirms the lack of a significant difference, and the fraction of systems with $\log_{10}(M_{\mathrm{cluster}}/M_\odot) > 14.7$ is identical ($18\%$) in the two samples.

Overall, radio-loud BCGs are not distinguished by differences in star formation or host cluster mass, but exhibit modestly elevated accretion rates within the same sub-Eddington regime. This suggests that radio-loud AGN activity reflects variations in gas supply and accretion within the ICM, rather than being driven by host-galaxy properties alone.

\subsection{Cluster context}

Brightest cluster galaxies reside at the centres of the most massive dark matter haloes in the Universe, where the thermodynamic state of the ICM regulates gas cooling and AGN fuelling \citep[e.g.][]{Cavagnolo2008,Hudson2010,Voit2015}. Having characterised the accretion properties of the BEAMS sample (Sect.~\ref{sec:results_accretion}) and shown that radio activity is not driven solely by star formation, we next examined whether these properties depend on the larger-scale cluster environment.

As a baseline, we first verified that BCG stellar mass shows little variation with cluster mass.\footnote{The corresponding figure is omitted for brevity.} The clusters span $14.2 \lesssim \log_{10}(M_{\mathrm{cluster}}/M_\odot) \lesssim 15.1$, consistent with the SZ-selected nature of the parent catalogue. Across this interval, BCG stellar masses remain uniformly high, with $\log(M_\star/M_\odot) \gtrsim 11$. A Spearman rank test yields $\rho = 0.12$ with $p = 0.18$, indicating no statistically significant correlation between stellar mass and total cluster mass. The intrinsic scatter in $\log(M_\star)$ at fixed cluster mass ($\sim0.2$--0.3 dex) exceeds any systematic shift in the median across the cluster-mass range probed. This behaviour is consistent with previous studies showing a relatively shallow dependence of BCG stellar mass on halo mass at the high-mass end \citep{DeLucia2007,Kravtsov2018}.

\begin{figure}
    \centering
    \includegraphics[width=0.8\columnwidth]{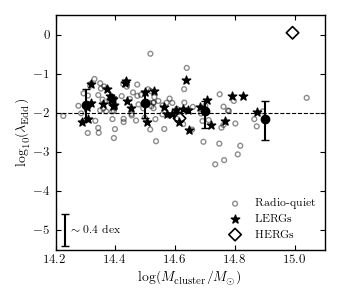}
\caption{Eddington-scaled accretion rate versus host cluster mass for the BEAMS BCGs. Grey open circles show radio-undetected BCGs, black stars radio-detected LERGs, and open diamonds HERGs. Filled circles show median $\log_{10}(\lambda_{\mathrm{Edd}})$ values in cluster-mass bins, with the 16th--84th percentile range. The dashed line marks $\lambda_{\mathrm{Edd}} = 10^{-2}$.}
    \label{fig:acc_rate_mcluster}
\end{figure}

\subsubsection{Accretion efficiency versus cluster mass}

We next examined whether Eddington-scaled accretion rates depend on cluster mass. Figure~\ref{fig:acc_rate_mcluster} shows $\log_{10}(\lambda_{\mathrm{Edd}})$ as a function of cluster mass. Across $14.2 \lesssim \log_{10}(M_{\mathrm{cluster}}/M_\odot) \lesssim 15.1$, $\lambda_{\mathrm{Edd}}$ exhibits a weak anti-correlation with cluster mass. A Spearman rank test yields $\rho = -0.25$ with $p = 0.0037$, indicating that BCGs in more massive clusters tend to accrete at slightly lower Eddington ratios on average. Restricting the analysis to the 34 radio-loud BCG AGNs with robust SED fits yields a consistent trend ($\rho = -0.25$), although it is not statistically significant ($p = 0.15$), reflecting the smaller sample size.
However, the intrinsic scatter in $\log_{10}(\lambda_{\mathrm{Edd}})$ ($\sim0.8$--1 dex) substantially exceeds the variation in the median accretion rate throughout the cluster-mass range. Combined with systematic uncertainties of the order of $\sim0.4$ dex, this implies that the observed trend is weak and should be interpreted with caution. In practice, total cluster mass does not uniquely determine the instantaneous accretion efficiency of BCGs. Consistent with this, we find no correlation between radio luminosity and cluster mass for radio-loud BCG AGNs (Spearman $\rho = -0.02$, $p = 0.90$), indicating that jet power is not determined by cluster mass.
By comparison, the correlation between $\lambda_{\mathrm{Edd}}$ and redshift (Fig.~\ref{fig:acc_rate_z}) is stronger, suggesting that cosmic epoch plays a more important role in shaping the typical accretion state of BCGs than cluster mass. This supports a picture in which AGN fuelling is governed primarily by local gas conditions within the ICM rather than by the overall mass of the cluster.

\begin{figure}
    \centering
    \includegraphics[width=0.92\columnwidth]{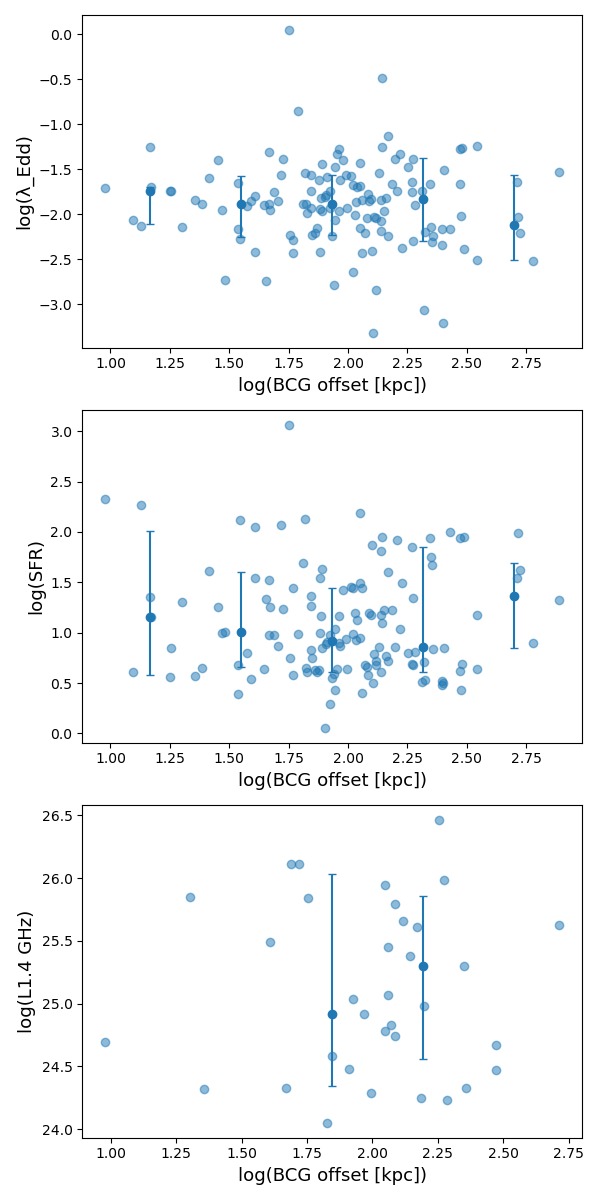}
    \caption{AGN and host-galaxy properties versus projected BCG--cluster offset.
    \textit{From top to bottom}: Eddington-scaled accretion rate, SFR, and radio luminosity.
    Points show individual BCGs; filled circles show the median values in offset bins, with the 16th--84th percentile range.
    No significant trends are observed in any panel.
    }
    \label{fig:offset_properties}
\end{figure}

\subsubsection{Dynamical state: BCG-cluster offsets}

The projected offset between the BCG and the cluster centre provides a useful probe of cluster dynamical state. In relaxed systems, the BCG is expected to reside close to the minimum of the gravitational potential, whereas larger offsets are generally associated with disturbed or merging clusters \citep[e.g.][]{Sanderson2009,Rossetti2016,Lopes2018}. This diagnostic is particularly relevant for SZ-selected samples such as BEAMS, which are expected to contain a larger fraction of dynamically unrelaxed systems. Mergers can temporarily enhance the SZ signal through shocks and pressure perturbations, potentially increasing the fraction of disturbed systems identified in SZ surveys \citep[e.g.][]{Rossetti2016,Cui2016}.

If the large-scale dynamical state of clusters plays a primary role in regulating AGN activity or star formation in BCGs, measurable trends with BCG--cluster separation might be expected. To test this, we examined the dependence of AGN and host-galaxy properties on projected offset, defined as the physical separation in kpc between the optical position of the BCG and the ACT cluster centroid (RA, Dec) reported by \citet{Hilton2021}. Figure~\ref{fig:offset_properties} shows the Eddington-scaled accretion rate, SFR, and radio luminosity as a function of projected offset in logarithmic space.
We find no statistically significant correlations between projected offset and any of the quantities considered. The Eddington ratio shows no dependence on offset ($\rho = -0.06$, $p = 0.51$), indicating no measurable connection between AGN accretion properties and cluster dynamical state within the uncertainties of our data. Similarly, the SFR exhibits no trend with offset ($\rho = -0.01$, $p = 0.90$), while the radio luminosity of the radio-detected subsample is likewise uncorrelated with offset ($\rho = -0.03$, $p = 0.87$), although the smaller sample size limits the statistical power of this test. The absence of trends is further supported by the flat behaviour of the binned medians across the full range of projected separations probed. Projected offsets provide only a lower limit on the true three-dimensional separation between the BCG and the cluster centre, and SZ-derived centroids may introduce additional uncertainty compared with X-ray-based definitions. These effects are expected to increase scatter rather than erase a strong intrinsic correlation, reinforcing the robustness of the null result.

These findings can be compared with previous studies linking BCG properties to cluster centring. For example, \citet{Sanderson2009} and \citet{Pasini2019} find connections between cooling, AGN activity, and BCG position, while \citet{Liu2025} report that the fraction of star-forming BCGs increases with offset, with star-forming systems preferentially found in dynamically disturbed clusters. Such trends are commonly interpreted as a consequence of merger-driven gas redistribution and enhanced galaxy interactions.
Our results are not inconsistent with this picture. While previous studies primarily probe the incidence of AGNs or star-forming activity, we instead considered continuous measures such as the SFR and accretion rate. The absence of correlations therefore suggests that, although cluster dynamical state may influence whether activity is triggered, it does not strongly regulate the level of activity once established. This interpretation is consistent with a scenario in which large-scale disturbances modulate gas supply to the cluster core, while the efficiency of black hole fuelling and star formation is governed primarily by local thermodynamic conditions and self-regulated feedback cycles \citep[e.g.][]{Gaspari2018,Fujita2020}.

A related consideration concerns the connection between radio power and cluster-scale properties. While correlations between radio power and cluster mass are well established for diffuse radio emission such as mini-haloes \citep[e.g.][]{Kolokythas2025}, studies of BCG-hosted radio AGNs generally find large intrinsic scatter and weak or absent correlations with cluster mass \citep[e.g.][]{Laferriere2020}. Our results are consistent with this picture, indicating that the instantaneous radio power of BCG AGNs is not strongly regulated by the global cluster potential.

Taken together, these results suggest that global environmental properties play at most a secondary role in regulating AGN activity in BCGs. While cluster-scale processes may influence gas transport towards the core, the accretion state of the central AGN is likely governed primarily by local thermodynamic conditions, such as cooling, entropy, and cold-gas availability \citep{Gaspari2013,HlavacekLarrondo2012}. Across $0.3 < z < 0.8$, the BCG population remains predominantly in the low-Eddington-ratio regime characteristic of maintenance-mode feedback.

\section{Discussion}
\label{sec:discussion}

\subsection{Maintenance-mode AGNs in BCGs}

Our results show that AGN activity in BCGs at $0.3 < z < 0.8$ is overwhelmingly dominated by LERGs, which comprise $\sim95\%$ of the radio-loud AGN population, with only two HERGs identified in the sample. The inferred Eddington ratios span the low to intermediate regime and are consistent with predominantly radiatively inefficient accretion.

These results indicate that kinetic, or maintenance-mode, feedback remains the dominant mode of black hole growth in cluster centrals over this redshift range. Radiatively efficient episodes do occur, but they appear to be rare and transient rather than representative of the typical accretion state. The persistence of this behaviour to $z\sim0.8$ suggests that maintenance-mode AGNs were already widespread in massive cluster galaxies by this epoch, consistent with the emergence of self-regulated AGN feedback in cluster cores \citep[e.g.][]{Fabian2012,McNamaraNulsen2012,Calzadilla2024}.

\subsection{What regulates AGN activity in BCGs?}

A key result of this work is that global cluster properties do not strongly regulate AGN activity in BCGs. We find, at most, weak trends between accretion rate and cluster mass, with no statistically significant dependence on dynamical state within the uncertainties of the present sample. This result holds despite the inclusion of both relaxed and dynamically disturbed systems. Such behaviour is expected for SZ-selected samples, which are less biased towards relaxed, cool-core clusters than X-ray surveys and therefore probe a more representative range of cluster dynamical states.
By contrast, accretion efficiency increases with redshift, indicating that fuel availability may play a more important role than global environment in regulating the instantaneous accretion state. This result is consistent with the increasing fraction of AGN-hosting BCGs reported in SZ-selected cluster samples at higher redshifts \citep{Somboonpanyakul2022} and with the broader evolution of gas fractions and cooling rates in massive haloes since $z\sim1$ \citep[e.g.][]{MadauDickinson2014}. The trend remains significant after excluding Phoenix\,A, indicating that it is characteristic of the broader BCG population rather than driven by a single extreme system.

These results point towards a scenario in which AGN fuelling is governed primarily by local thermodynamic conditions in cluster cores, such as cooling time, entropy, and the balance between heating and precipitation \citep[e.g.][]{Churazov2005,Voit2015}. In this framework, global halo properties define the environment in which BCGs reside but do not uniquely determine the level of accretion or jet power at a given time.

\subsection{BCGs in the context of the radio-loud AGN population}

Comparison with field radio-loud AGNs shows that BCG-hosted LERGs are not intrinsically distinct in their radio properties. At fixed redshift, their radio luminosities are consistent with those of field LERGs once the luminosity limits imposed by the RACS selection function are taken into account. The primary difference lies in host-galaxy mass: BCG radio-loud AGNs occupy the extreme high-mass end of the LERG population. This supports the well-established connection between stellar mass and the probability of hosting a radio-loud AGN \citep[e.g.][]{Best2023,Kondapally2025}.

These findings suggest that the physics of radio-mode accretion and jet production is broadly similar across environments. The role of the cluster environment is therefore not to fundamentally alter the nature of AGN activity, but to provide the conditions under which the most massive galaxies can sustain long-lived, maintenance-mode feedback.

\subsection{Implications for feedback and gas regulation}

The absence of strong correlations between AGN properties and global cluster parameters, combined with the weak link between radio power and radiative accretion efficiency, has important implications for models of AGN feedback. In particular, the results favour self-regulated feedback scenarios in which gas cooling and AGN heating are coupled through local processes in the cluster core.
In such models, cooling instabilities lead to the condensation of cold gas, which fuels the central black hole and triggers jet activity. The resulting mechanical feedback then reheats the surrounding medium, maintaining a long-term balance between heating and cooling \citep[e.g.][]{Gaspari2013,Gaspari2017,Voit2015}.

The broad range of Eddington ratios and the lack of a tight connection between jet power and radiative output are natural outcomes of this framework, reflecting the stochastic and multiphase nature of gas accretion in cluster cores. Although the absolute SFRs of passive systems remain uncertain in the absence of FIR constraints, the observed relative trends with redshift are unlikely to be driven solely by SED-fitting systematics.

\subsection{Caveats}

Several limitations should be noted. The RACS flux limit restricts the analysis to moderate- and high-luminosity radio sources, potentially missing low-power radio AGNs at higher redshifts. The measured radio-loud BCG fraction of 22\% should therefore be regarded as a lower limit. Deeper MeerKAT observations recover an additional 15 radio-detected BCGs below the RACS threshold, indicating that incompleteness mainly affects the low-luminosity population.

The absence of FIR constraints limits the accuracy of the dust luminosity and obscured SFR, introducing additional systematic uncertainty in the absolute SFR normalisation and in the inferred positions of individual BCGs relative to the star-forming main sequence. AllWISE profile-fit photometry may also introduce additional uncertainties for the most extended low-redshift BCGs. The \textsc{X-CIGALE} AGN component assumes a radiatively efficient AGN template, so the derived bolometric luminosities and Eddington-scaled accretion rates should be regarded as upper limits for LERGs.

Projected BCG--cluster offsets introduce scatter through projection effects and centroid uncertainties, while the SZ selection may still retain biases related to cluster morphology and dynamical state. The ACT sample also spans a relatively narrow halo-mass range and is dominated by massive clusters, limiting our ability to investigate environmental trends at lower halo masses.
Despite these limitations, the uniform selection and multiwavelength characterisation of the BEAMS sample provide a representative view of AGN activity in BCGs at intermediate redshifts.

\section{Conclusions}
\label{sec:conclusion}

We investigated 171 BCGs in SZ-selected clusters over $0.3 < z < 0.8$ as part of the BEAMS programme, combining SALT spectroscopy, WISE MIR data, RACS radio observations, and \textsc{X-CIGALE} SED fitting. This provides a uniform characterisation of AGN incidence, accretion state, and host-galaxy properties in an approximately mass-limited cluster sample. Our main findings are:
\begin{enumerate}

\item Brightest cluster galaxy radio-loud AGNs are dominated by maintenance-mode accretion.
Radio-loud AGNs are overwhelmingly LERGs, comprising $\sim95\%$ of the radio-loud population, with only two HERGs identified in the sample. The inferred Eddington ratios span $-3 \lesssim \log_{10}\lambda_{\mathrm{Edd}} \lesssim -1$ and are consistent with predominantly radiatively inefficient accretion and kinetic-mode feedback.

\item Accretion efficiency evolves with redshift but shows only a weak dependence on global environment.
The Eddington-scaled accretion rate increases towards higher redshifts, while its dependence on cluster mass is weak and no statistically significant dependence is found on cluster dynamical state within the uncertainties of the present sample. This suggests that the evolution of gas availability may play a more important role than global cluster properties in regulating AGN activity.

\item Brightest cluster galaxy radio-loud AGNs occupy the high-mass end of the LERG population.
At fixed redshift, BCG radio sources are consistent in radio power with field LERGs once selection effects are taken into account, but are hosted by galaxies that are more massive by $\sim0.8$ dex. These results indicate that BCG radio-loud AGNs form a continuous extension of the broader LERG population rather than a clearly distinct population.

\item Jet power is not tightly coupled to star formation or radiative accretion.
The apparent correlation between radio luminosity and SFR is largely driven by their shared redshift evolution, with no compelling evidence for a direct link between jet power and either SFR or $\lambda_{\mathrm{Edd}}$.

\end{enumerate}

Together, these results indicate that AGN activity in BCGs is governed primarily by local thermodynamic conditions in cluster cores rather than global cluster properties. The BCG radio-loud AGN population at $0.3 < z < 0.8$ is dominated by radiatively inefficient, kinetic-mode systems, with evidence for increasing accretion efficiency towards higher redshifts. While the relative importance of cooling- and merger-driven fuelling may evolve with redshift, maintenance-mode feedback was already well established in massive cluster galaxies by $z\sim1$.

Future work combining deeper radio observations with high-resolution X-ray imaging of cluster cores, together with ALMA follow-up where available, will enable more direct tests of the connection between gas cooling, cold-gas reservoirs, black hole accretion, and AGN feedback across cosmic time in massive clusters. In particular, jointly linking X-ray morphological indicators, molecular gas content, and radio AGN properties will provide a more physically motivated framework for interpreting the observed Eddington ratio distribution and its relation to BCG--cluster offset. Extending such analyses to lower-mass haloes beyond the present SZ-selected sample will be essential for determining how AGN fuelling and feedback depend on environment across the full range of group and cluster systems.

\section*{Data availability}

The catalogue of the 160 BCGs analysed in this work, together with their
principal properties, is available in electronic form at the CDS via
anonymous ftp to \url{cdsarc.u-strasbg.fr} (130.79.128.5) or via
\url{https://cdsweb.u-strasbg.fr/cgi-bin/qcat?J/A+A/}.

\begin{acknowledgements}
Some of the observations reported in this paper were obtained with the Southern African Large Telescope (SALT). M. Hilton acknowledges support from the National Research Foundation of South Africa (grant numbers 137975, 97792). J. Delhaize acknowledges partial research support by the National Research Foundation of South Africa (Ref Number CSUR240426216203). We are grateful to the late Prof.~T.~H.~Jarrett for providing the $k$-corrected WISE colour catalogue  used in this work. The authors thank Dr. Imogen Whittam for providing the MIGHTEE-COSMOS catalogue  and the associated \textsc{LePhare}
SED fitting results used in this work. We acknowledge the use of the ilifu cloud computing facility – \url{www.ilifu.ac.za}, a partnership between the University of Cape Town, the University of the Western Cape, Stellenbosch University, Sol Plaatje University and the Cape Peninsula University of Technology. The ilifu facility is supported by contributions from the Inter-University Institute for Data Intensive Astronomy (IDIA – a partnership between the University of Cape Town, the University of Pretoria and the University of the Western Cape), the Computational Biology division at UCT and the Data Intensive Research Initiative of South Africa (DIRISA).

\end{acknowledgements}

\bibliography{manuscript}{}
\bibliographystyle{aa}

\begin{appendix}

\section{Supplementary material}
\label{appendix:agn-class}
\begin{figure}[H]
    \centering
    \includegraphics[width=0.9\columnwidth]{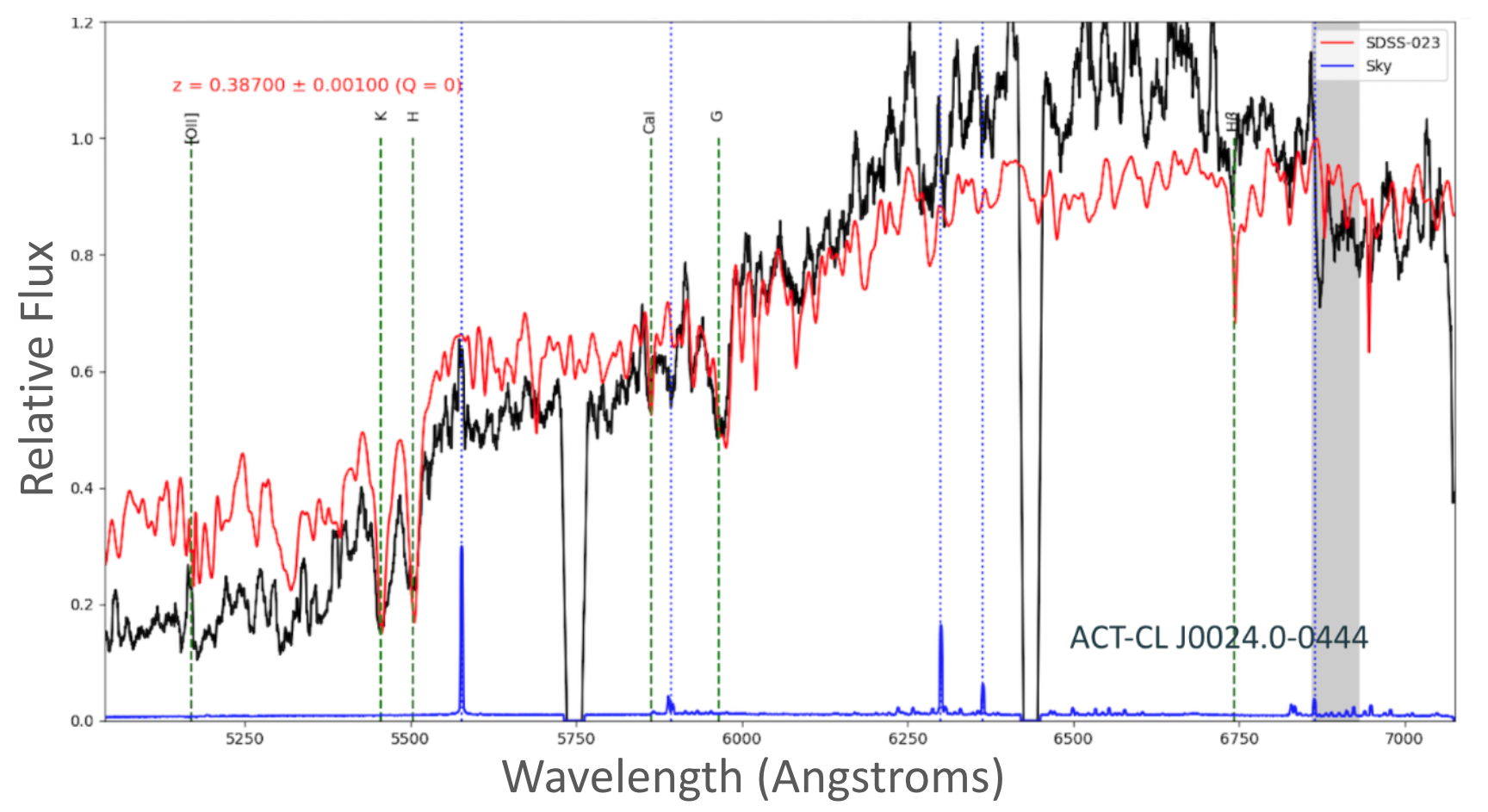}
\caption{Example one-dimensional SALT RSS spectrum of an absorption-line BCG at $z=0.387$, reduced with the \texttt{RSSMOSPipeline}. Prominent absorption features used for the redshift determination, including Ca H and K, are labelled.}
    \label{fig:bcg_example_abs}
\end{figure}

\begin{figure}[H]
    \centering
    \includegraphics[width=0.7\columnwidth]{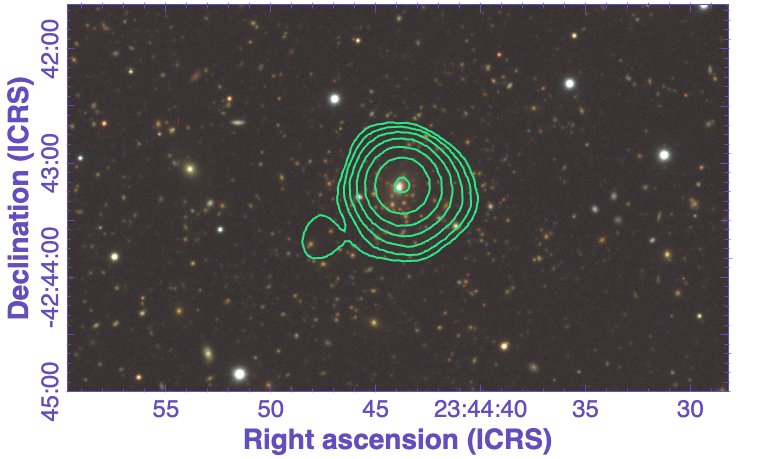}
\caption{DES  $gri$ red-green-blue image of the Phoenix cluster BCG overlaid with RACS radio continuum contours. Contours are drawn at 3, 5, 9, 17, 33, 65, and 129$\sigma$, where $\sigma = 4\times10^{-4}\,\mathrm{Jy\,beam^{-1}}$.}
    \label{DES_racs}
\end{figure}

\begin{figure}[H]
    \centering
    \includegraphics[width=0.85\columnwidth]{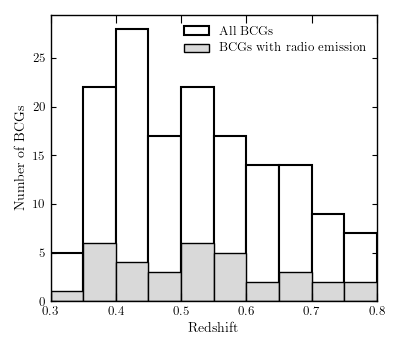}
\caption{Redshift distribution of the full BEAMS BCG sample (outlined histogram) and the RACS-detected subset (filled histogram).}
    \label{fig:redshift_dist}
\end{figure}

\begin{figure}[H]
    \centering
    \includegraphics[width=\columnwidth]{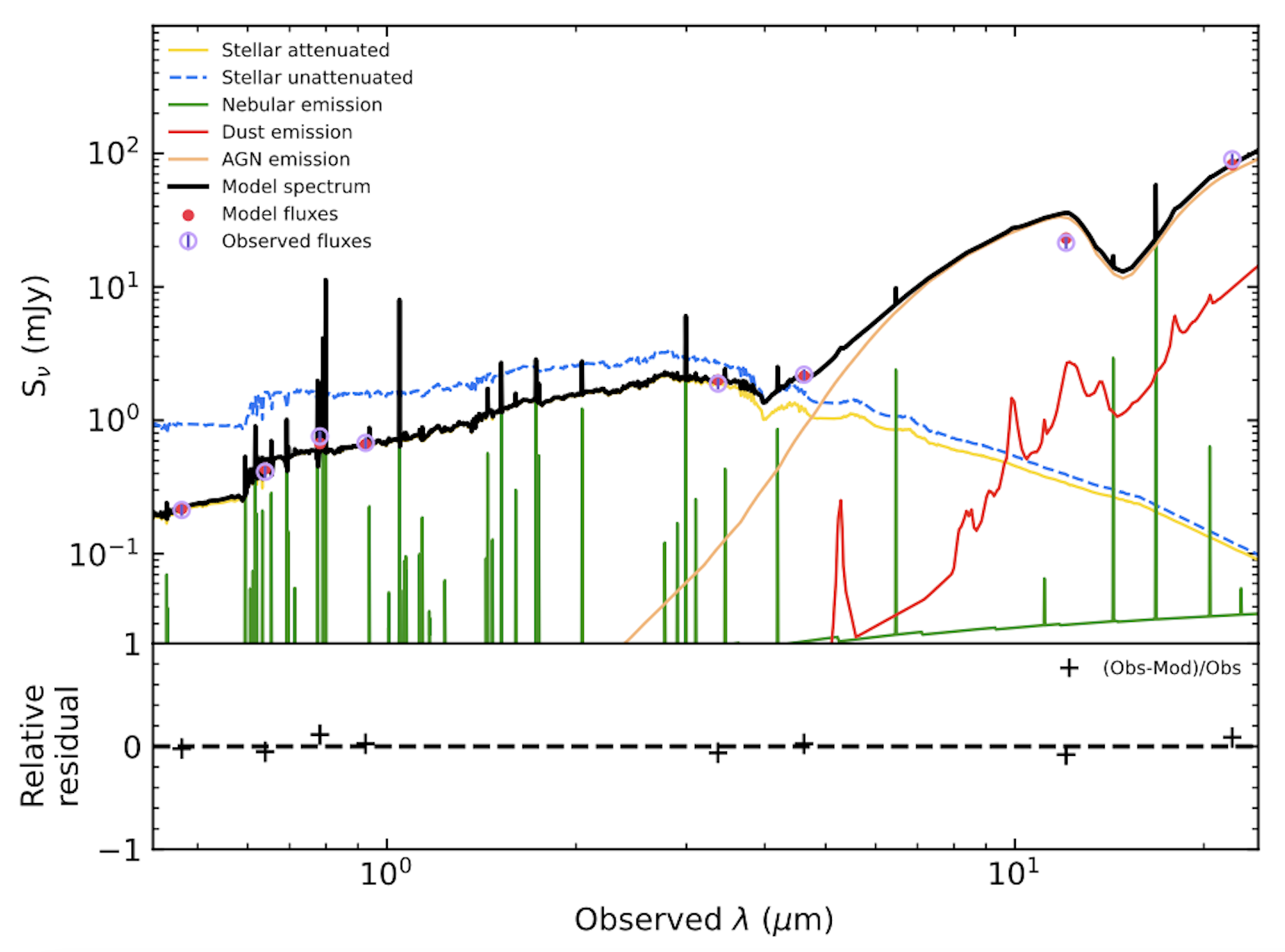}
\caption{\textit{Top}: Example \textsc{X-CIGALE} SED fit for the Phoenix cluster BCG. Symbols show the observed fluxes and corresponding model fluxes. The best-fit SED (black) is decomposed into stellar, nebular, dust, and AGN components, with the stellar component including attenuated and unattenuated emission. \textit{Bottom}: Relative residuals ($\chi^2_{\rm red}=0.25$).}
    \label{cigale_examp}
\end{figure}

\begin{figure}[H]
    \centering
    \includegraphics[width=\columnwidth]{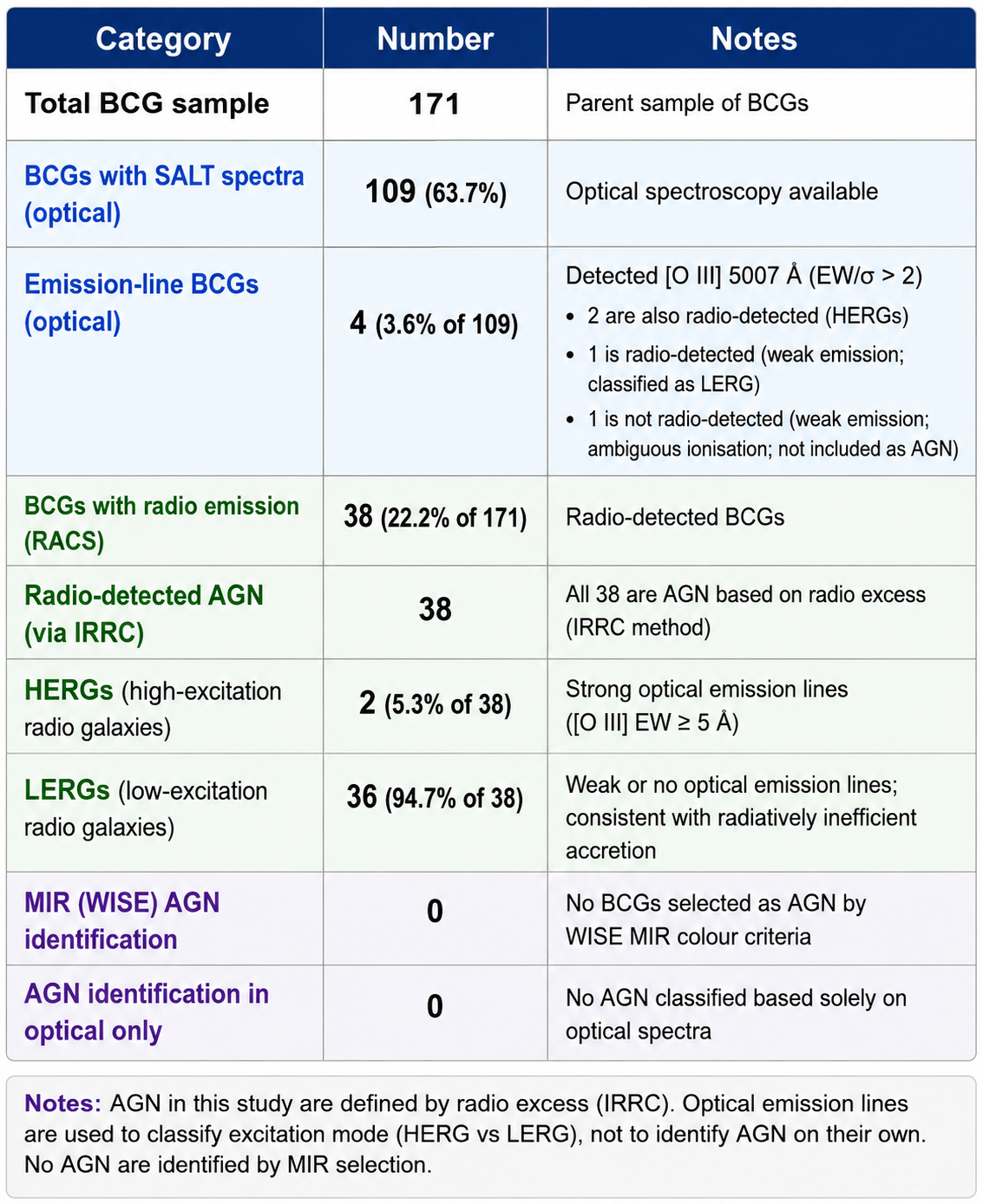}
    \caption{Summary of the BCG sample and AGN classification.}
    \label{fig:agn_class}
\end{figure}

\noindent
This appendix provides supplementary material supporting the analysis presented in the main text. We include an example reduced SALT RSS spectrum used for the spectroscopic redshift determination (Fig.~\ref{fig:bcg_example_abs}), an illustration of the radio morphology of the Phoenix cluster BCG (Fig.~\ref{DES_racs}), the redshift distributions of the full BEAMS sample and the RACS-detected subset (Fig.~\ref{fig:redshift_dist}), an example \textsc{X-CIGALE} SED fit (Fig.~\ref{cigale_examp}), and a summary of the AGN classifications adopted in this work (Fig.~\ref{fig:agn_class}).

\FloatBarrier
\section{SED fitting uncertainties and reliability}
\label{appendix:SED-uncer}

The uncertainties on the derived parameters were obtained from the Bayesian posterior distributions computed by \textsc{X-CIGALE}, which reflect both the propagation of photometric errors and intrinsic model degeneracies. The likelihood-weighted means correspond to the posterior expectation values, while the 16th--84th percentile ranges define the $1\sigma$ confidence intervals.

For the 133~BCGs with consistent SED fits, the $1\sigma$ uncertainties in $\log(M_{\star}/M_{\odot})$ range from 0.03 to 0.17\,dex, with a median of $\simeq0.13$\,dex, showing that stellar masses are tightly constrained. Uncertainties in $\log(\mathrm{SFR}/M_{\odot}\,\mathrm{yr}^{-1})$ span 0.08--1.3\,dex, with a median of $\simeq0.4$\,dex. These uncertainties reflect the limited wavelength coverage (DES$+$AllWISE), the degeneracy between low SFRs and evolved stellar populations, and the relatively low S/N of the WISE $W3$ and $W4$ bands for many passive BCGs. In such cases, the weak MIR constraints primarily broaden the posterior distributions of dust luminosity and SFR rather than introducing strong systematic biases. These ranges are consistent with previous \textsc{CIGALE} and \textsc{X-CIGALE} applications \citep[e.g.][]{Ciesla2015,Buat2021,Masoura2018,Mountrichas2022a}.

To assess the robustness of the recovered parameters, we analysed the mock catalogues produced by \textsc{X-CIGALE}, in which the best-fit SEDs were perturbed according to the photometric uncertainties and refitted using the same model grid. Comparison between the input and recovered parameters shows excellent agreement. For stellar masses, $\langle\Delta\log M_{\star}\rangle = 0.011$\,dex with $\sigma_{\log M_{\star}} = 0.085$\,dex; for SFRs, $\langle\Delta\log \mathrm{SFR}\rangle = 0.021$\,dex with $\sigma_{\log \mathrm{SFR}} = 0.30$\,dex; and for the AGN fractional contribution, $\langle\Delta\log f_{\mathrm{AGN}}\rangle = -0.065$\,dex with $\sigma_{\log f_{\mathrm{AGN}}} = 0.32$\,dex. The small offsets indicate negligible systematic bias within the adopted model framework, while the broader scatter in SFR and $f_{\mathrm{AGN}}$ reflects the limited MIR constraints and the known coupling between dust heating by star formation and AGN emission. These uncertainties therefore primarily represent statistical errors arising from photometric uncertainties, given that the underlying model assumptions are held fixed.

Systematic modelling effects are not captured by the mock analysis and contribute additional uncertainty. A recent intercomparison of 14~SED-fitting codes by \citet{Pacifici2023} found that variations in model assumptions alone introduce typical dispersions of $\simeq0.1$\,dex in $\log M_{\star}$ and $\simeq0.3$\,dex in $\log\mathrm{SFR}$. These values should be interpreted as an approximate systematic floor, which, when combined with the statistical uncertainties derived above, defines the total uncertainty budget of the SED-derived parameters.

Taken together, the good fit statistics, internal consistency, and mock-recovery validation indicate that the derived stellar masses and AGN fractions are robust within the assumptions of the adopted SED-fitting framework for the 133~BCGs analysed in the following sections. Given the lack of FIR constraints, the SFRs, particularly their absolute normalisation, should be interpreted with caution, as the obscured star-formation component is subject to additional systematic uncertainty.

\section{Comparison of black hole mass estimators}
\label{appendix:mbh_comparison}

In the main analysis, black hole masses for the BEAMS BCGs are estimated from host stellar masses using the empirical relation
\begin{equation}
M_{\mathrm{BH}} = 0.0014\,M_\star,
\end{equation}
derived by \citet{Haering2004}. This relation provides a practical and internally consistent estimate of $M_{\mathrm{BH}}$ for the full BCG sample.

To assess the robustness of this choice, we performed an independent estimate of $M_{\mathrm{BH}}$ based on the stellar velocity dispersion, $\sigma_\star$, using the well-established $M_{\mathrm{BH}}$--$\sigma_\star$ relation of \citet{Tremaine2002},
\begin{equation}
\log\left(\frac{M_{\mathrm{BH}}}{M_\odot}\right) =
8.13 + 4.02 \log\left(\frac{\sigma_\star}{200~\mathrm{km\,s^{-1}}}\right).
\end{equation}

Measurements of $\sigma_\star$ for the BEAMS BCG sample are taken from \citet{Loubser2025}, where velocity dispersions were derived from stacked optical spectra in redshift bins. These measurements therefore do not provide object-by-object black hole masses and are available only over a restricted redshift range that does not extend to $z\simeq0.8$. For this reason, the $\sigma_\star$-based black hole masses are not used in the main analysis but instead serve as an independent consistency check.

\begin{figure}
    \centering
    \includegraphics[width=0.83\columnwidth]{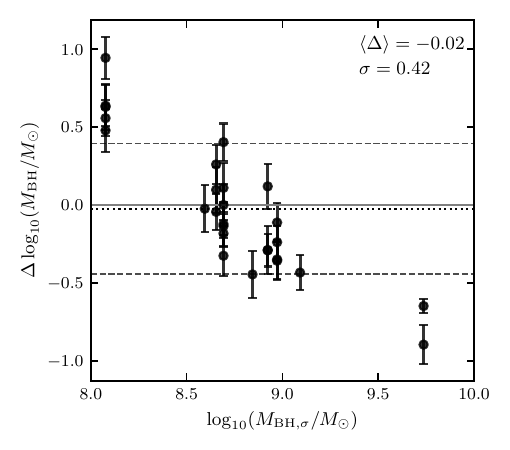}
\caption{Comparison of black hole masses estimated from the $M_{\mathrm{BH}}$--$\sigma_\star$ relation of \citet{Tremaine2002} and the stellar-mass scaling relation of \citet{Haering2004}. The dotted horizontal line marks the mean offset, while the dashed lines show $\pm1\sigma$. Error bars show the propagated uncertainties in the stellar-mass-based black hole masses.}
    \label{fig:mbh_compare}
\end{figure}

Figure~\ref{fig:mbh_compare} compares the two black hole mass estimates. We find a negligible mean offset of $\langle \Delta \log_{10} M_{\mathrm{BH}} \rangle = -0.023$ dex, indicating no systematic bias between the methods. The observed dispersion is $\sigma = 0.419$ dex, which is consistent with expectations. The stellar-mass-based relation of \citet{Haering2004} exhibits an intrinsic scatter of $\sim0.3$ dex, and the $M_{\mathrm{BH}}$--$\sigma_\star$ relation likewise has an intrinsic dispersion of order $\sim0.3$ dex. The quadrature combination of these independent uncertainties, $\sqrt{0.3^2 + 0.3^2} \approx 0.42$ dex, closely matches the observed dispersion. The agreement therefore indicates that the two estimators are statistically consistent within their expected intrinsic scatter. Other calibrations of the $M_{\mathrm{BH}}$--$\sigma_\star$ relation, particularly those tailored to massive early-type galaxies and BCGs, have also been proposed \citep[e.g.][]{McConnellMa2013}. Adopting such relations would primarily modify the absolute normalisation of the inferred black hole masses but would not alter the qualitative agreement demonstrated here.
Overall, this comparison confirms that the stellar-mass-based black hole masses adopted throughout the main body of this work are robust and appropriate for characterising the accretion properties of BCGs in the BEAMS sample.

\end{appendix}

\end{document}